\documentclass[sigconf]{aamas} 

\usepackage{balance} 

\setcopyright{ifaamas}
\acmConference[AAMAS '22]{Proc.\@ of the 21st International Conference
on Autonomous Agents and Multiagent Systems (AAMAS 2022)}{May 9--13, 2022}
{Online}{P.~Faliszewski, V.~Mascardi, C.~Pelachaud,
M.E.~Taylor (eds.)}
\copyrightyear{2022}
\acmYear{2022}
\acmDOI{}
\acmPrice{}
\acmISBN{}

\usepackage{wrapfig}
\usepackage{graphicx} 
\usepackage{graphicx}  

\usepackage{graphicx}  
\usepackage{hyperref}

\usepackage{booktabs}
\usepackage{macros}

\title{Controller Synthesis for Omega-Regular and Steady-State Specifications} 

\author{Alvaro Velasquez}
\affiliation{
  \institution{Air Force Research Laboratory}
  \country{Rome, New York, USA}}
\email{alvaro.velasquez.1@us.af.mil}

\author{Ismail Alkhouri}
\affiliation{
  \institution{University of Central Florida}
  \country{Orlando, Florida, USA}
}
\email{ialkhouri@knights.ucf.edu}

\author{Andre Beckus}
\affiliation{
  \institution{Air Force Research Laboratory}
  \country{Rome, New York, USA}}
\email{andre.beckus@us.af.mil}

\author{Ashutosh Trivedi}
\affiliation{
  \institution{University of Colorado Boulder}
  \country{Boulder, Colorado, USA}}
\email{ashutosh.trivedi@colorado.edu}

\author{George Atia}
\affiliation{
  \institution{University of Central Florida}
  \country{Orlando, Florida, USA}
}
\email{george.atia@ucf.edu}

\acmSubmissionID{670}

\begin{abstract}
Given a Markov decision process (MDP) and a linear-time ($\omega$-regular or Linear Temporal Logic) specification which reasons about the infinite-trace behavior of a system, the controller synthesis problem aims to compute the optimal policy that satisfies said specification. Recently, problems that reason over the complementary infinite-frequency behavior of systems have been proposed through the lens of steady-state planning or steady-state policy synthesis. This entails finding a control policy for an MDP such that the Markov chain induced by the solution policy satisfies a given set of constraints on its steady-state distribution. This paper studies a generalization of the controller synthesis problem for a linear-time specification under steady-state constraints on the asymptotic behavior of the agent. We present an algorithm to find a deterministic policy satisfying
$\omega$-regular and steady-state constraints by characterizing the solutions as
an integer linear program, and experimentally evaluate our approach. 

\end{abstract}


\begin{document}

\maketitle



\section{Introduction}
\label{sec:intro}
The controller synthesis problem is often used to establish safety and performance guarantees of stochastic systems such as Markov decision processes (MDPs) by inducing Markov chains exhibiting some desirable behavior. The $\omega$-regular languages~\cite{Thomas90b,Perrin04} provide an expressive formalism to unambiguously express such safety and progress properties of MDPs, while Linear Temporal Logic (LTL) provides a convenient and interpretable way to encode such $\omega$-regular properties. For the verification or synthesis of systems subject to these properties, an $\omega$-regular objective is usually translated into a finite-state machine that monitors the traces of the MDP~\cite{deAlfa98}. Successful executions cause the finite-state machine to take certain accepting transitions infinitely often, and ultimately avoid certain rejecting transitions. That is, $\omega$-regular objectives reason about the asymptotic trace behavior of an MDP. The related notion of asymptotic frequency of states visited is not accounted for in such objectives. To illustrate the utility of being able to reason about both of these types of behavior, consider a simple robot tasked to explore terrain on Mars. For such a mission, one may come up with $\omega$-regular specifications that its traces of behavior should satisfy. For example, we may impose that, whenever a state labeled “ice” is encountered, the robot must collect a sample and drop it off at a state labeled “base”. Furthermore, the robot may also need to spend a certain proportion of its time -- but not too much time so as not to conflict with gathering ice samples – exploring certain regions of the Martian landscape. Indeed, the robot may be requested to spend at least 25\% of its time in regions of interest, but no more than 50\% of its time. This is easily encoded as a steady-state specification. Such specifications cannot be directly expressed in LTL.


LTL controller synthesis through probabilistic model checking approaches~\cite{deAlfa98,principlesOfModelChecking} generally begins by computing the product MDP from the original MDP and the finite-state machine representation of the given objective. Then, the union of accepting maximal end components (AMECs) are computed and a policy is found such that the agent reaches some such component. Once there, actions can be chosen arbitrarily such that all states within the AMEC are visited infinitely often, thereby ensuring that the acceptance condition of the automaton is met and the objective is therefore satisfied by said policy. Generally, this choice of actions within the AMECs is arbitrary. \textit{However, it is evident that these choices are critical in the situations with constraints on the steady-state distribution.}
This distribution characterizes the asymptotic frequency behavior of a Markov chain induced by some policy in an MDP. 

The controller synthesis problem subject to steady-state specifications has been explored recently \cite{SSC, ijcai2019-784, ijcai2020-563, atia2021steady} and the integration of LTL constraints has been considered for stochastic policy settings \cite{kvretinsky2021ltl}. In this paper, we seek to complement and unify much of the preceding work by reasoning about both $\omega$-regular properties as well as steady-state distributions
simultaneously and without making common assumptions of ergodicity on the underlying MDP. The proposed approach finds an optimal expected-reward deterministic policy that satisfies given $\omega$-regular and steady-state specifications. The computation of deterministic policies is an important avenue of inquiry when guarantees or predictable behavior are desired \cite{sarathy2021spotter}.

\section{Preliminaries}
\label{sec:prelims}
We recall classical definitions and introduce notation for the paper. 

\vspace{1em}
\noindent\textbf{Markov Decision Processes.}
A \emph{probability distribution} over a finite set $S$ is a function $d: S \rightarrow [0, 1]$ such that $\sum_{s \in S} d(s) = 1$. Let $D(S)$ denote the set of all discrete distributions over $S$. A Labeled Markov Decision Process (LMDP) $\mathcal{M}$ is a tuple $(S, \beta, A, T, R, AP, L)$, where $S$ is a finite set of states, $\beta \in D(S)$ is the initial state distribution, $A$ is a finite set of actions, $T: S \times A \to D(S)$ is the transition function, $R: S \times A \times S \rightarrow \Real$ is the reward signal, $AP$ is the set of atomic propositions, and $L:	S \rightarrow 2^{AP}$ is the \emph{labeling function}.
 

For any state $s \in S$, we let $A(s)$ denote the set of actions that can be selected in state $s$. For states $s, s' \in S$ and $a \in A(s)$, $T(s,a)(s')$ equals $p(s' | s, a)$. A {\it run} of $\mathcal{M}$ is a sequence $\seq{s_0, a_1, s_1, \ldots} \in S \times (A \times S)^*$ such that $p(s_{i+1} | s_{i}, a_{i+1}) > 0$ for all $i \geq 0$. A finite run is a finite such sequence. When convenient, runs are sometimes defined as sequences of states, without including actions. For a {\it run} $r = \seq{s_0, a_1, s_1, \ldots}$, we define the corresponding labeled run as $L(r) = \seq{L(s_0), L(s_1), \ldots} \in \big( 2^{AP} \big)^+$. A policy (or a strategy) is a recipe for a decision-maker to resolve the non-determinism of the LMDP. A policy in $\mathcal{M}$ is a function $\pi : S^+ \to D(A)$ mapping finite runs to actions. A policy is \emph{finite-memory} if it remembers a finite amount of information about the past and a finite-memory policy can be represented using a finite-state machine. In this paper, we are interested in finite-memory deterministic policies of the form $\pi: S \times Q \rightarrow A$, where $Q$ is a set of memory modes. 
This memory is obtained from the finite-state machine representation of the given linear-time specification to be satisfied. We write $\pi(a | s, q) \in \{0, 1\}$ for the probability of choosing action $a$ in the state $s$ when the memory mode is $q$. For the remainder of this paper, we assume finite-memory deterministic policies $\pi$. 
For an LMDP $\mathcal{M}  = (S, \beta, A, T, R, AP, L)$, a finite-memory deterministic policy $\pi$ resolves its non-determinism and gives rise to a Labeled Markov Chain (LMC) $\mathcal{M}_\pi = (S_\pi, \beta_\pi, T_\pi, R_\pi, AP_\pi, L_\pi)$.
Note that an LMC is an LMDP whose set of actions is a singleton and hence can be omitted. It is customary to represent the probabilistic
transition function $T$ of the LMC as a matrix such that $T_{i, j} = T(s_i)(s_j)$. When other information is not pertinent, we write an LMC as $(S, T)$.



Given an LMDP $\mathcal{M} = (S, \beta, A, T, R, AP, L)$, we define its underlying directed graph $G_\mathcal{M} = (V, E)$, where $V = S$ and $E \subseteq S \times S$ is such that $(s, s') \in E$ if $T(s, a)(s') > 0$ for some $a\in A(s)$. A sub-MDP of $\mathcal{M}$ is an LMDP $\mathcal{M}' = (S', \beta', A', T', R', AP', L')$, where $S' \subseteq S$, $A' \subseteq A$ is such that $A'(s) \subseteq A(s)$ for every $s \in S'$, and $\beta'$, $T'$, $R'$ and $L'$ are analogous to $\beta$, $T$, $R$, and $L$ when restricted to $S'$ and $A'$. An {\it end component}~\cite{deAlfa98} of an LMDP $\mathcal{M}$ is a sub-MDP $\mathcal{M}'$ of $\mathcal{M}$ such that $G_{\mathcal{M}'}$ is strongly connected. A {\em bottom strongly connected component} (BSCC) of an LMC is any of its maximal end components (MECs), where a MEC is an end component that is maximal under set inclusion.




\vspace{1em}
\noindent\textbf{Linear-Time Specifications.}
Given the set of atomic propositions $AP$ of an LMDP $\mathcal{M}$, a linear-time property of $\mathcal{M}$ is characterized by an $\omega$-language, i.e., a set of infinite sequences over the alphabet $\Sigma = 2^{AP}$. Formally, an $\omega$-\emph{word} $w$ on an alphabet $\Sigma$ is a function $w \colon \mathbb{N} \to \Sigma$.  We abbreviate $w(i)$ by $w_i$. The set of $\omega$-words on $\Sigma$ is written $\Sigma^\omega$ and a subset of $\Sigma^\omega$ is an $\omega$-\emph{language}. We are interested in expressing properties using $\omega$-regular languages given as a type of finite-state machine. In this context, we choose deterministic Rabin automata (DRA) as defined in the sequel.

    A \emph{deterministic Rabin automaton} (DRA) $\mathcal{A}$ is a tuple $(\Sigma, Q, q_0, \delta, F)$, where $\Sigma$ is a finite \emph{alphabet}, $Q$ is a finite set of \emph{states}, $q_0 \in Q$ is the \emph{initial state}, $\delta : Q \times \Sigma \to Q$ is the \emph{transition function}, and $F = \set{(B_i, G_i) \in 2^Q\times 2^Q}_{i \in [m]}$ is the \emph{Rabin acceptance condition}.  
A \emph{run} $r$ of a DRA $\mathcal{A}$ on $w \in \Sigma^\omega$ is an $\omega$-word
$r_0, w_0, r_1, w_1, \ldots$ in $(Q \cup \Sigma)^\omega$ such that $r_0 = q_0$
and, for $i > 0$, $r_i = \delta(r_{i-1},w_{i-1})$. We write $\infi(r) \subseteq
Q$ for the set of states that appear infinitely often in the run $r$. A run $r$
of a DRA $\mathcal{A}$ is \emph{accepting} if there is some $(B, G) \in F$ such that
$\infi(r) \cap B = \emptyset$ and $\infi(r) \cap G \neq \emptyset$. The
\emph{language} of $\mathcal{A}$ (or, \emph{accepted} by $\mathcal{A}$) is the subset of words
in $\Sigma^\omega$ that have accepting runs in $\mathcal{A}$. A language is
$\omega$-\emph{regular} iff it is accepted by a DRA \cite{principlesOfModelChecking}. 

\begin{figure}
\begin{center}
\includegraphics[width=1.05\columnwidth]{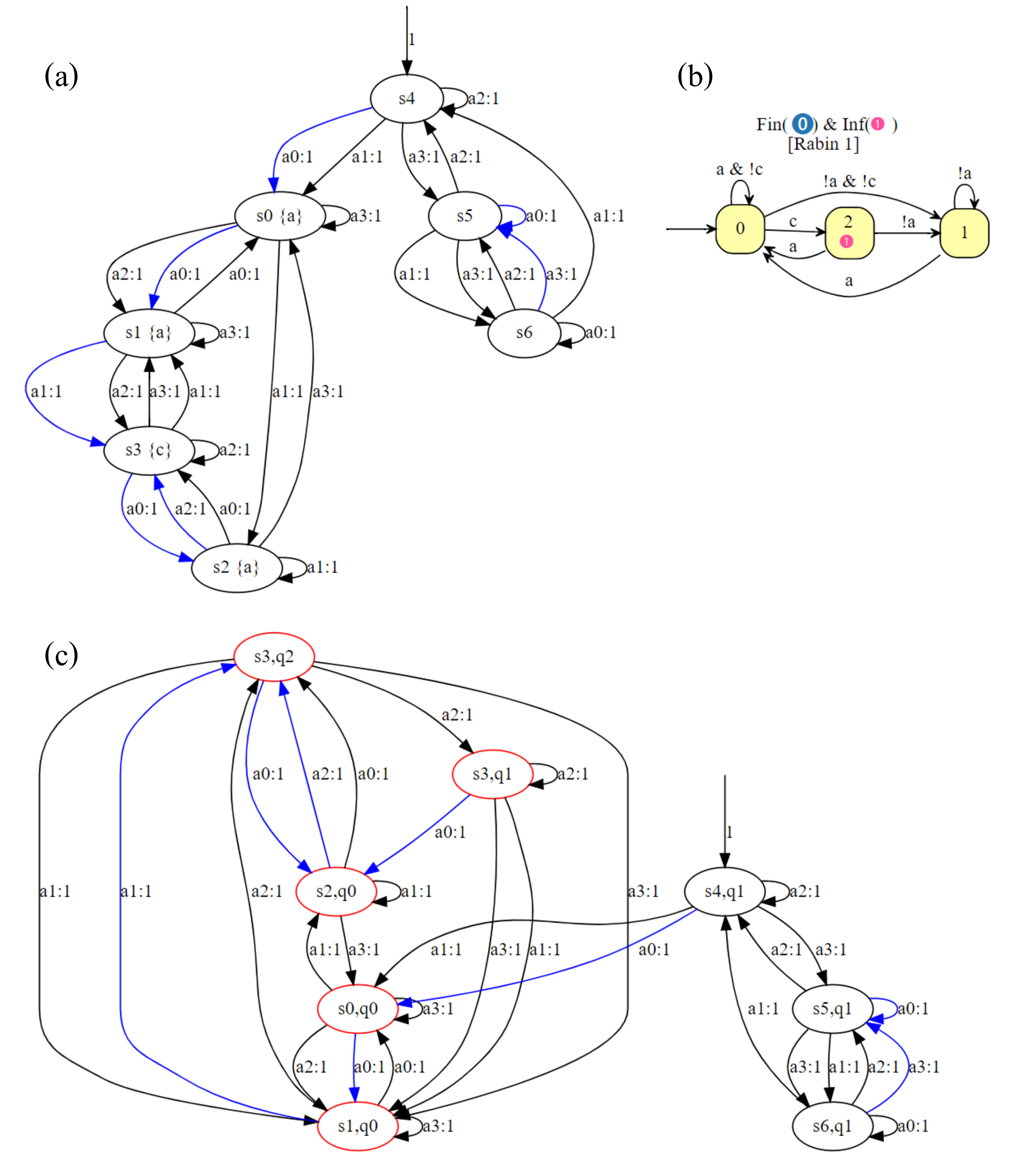}
\caption{(a) LMDP $\mathcal{M} = (S, \beta, A, T, R = \emptyset, AP, L)$, where $S = \{s_0, \dots, s_6\}$, $\beta(s_4) = 1$, $A = \{a_0, \dots, a_3\}$, $AP = \{a, c\}$, and $L(s_0) = L(s_1) = L(s_2) = \{a\}, L(s_3) = \{c\}$. The transition function is deterministic and shown in the figure by the transitions $a_i:1$ between states $s, s'$ denoting that $T(s, a_i)(s') = 1$. The blue transitions define a policy $\pi$ which induces a unichain LMC $\mathcal{M}_\pi$ (The isolated component consisting of states $s_5$ and $s_6$ is ignored since it is unreachable). (b) The DRA $\mathcal{A} = (Q, q_0, \Sigma, \delta, F = \{(B_i, G_i)\}_i)$ is shown, where $Q = \{q_0, q_1, q_2\}$, $\Sigma = 2^{AP}$, $F = \{(\emptyset, \{q_2\})\}$ and $!, \&$ denote logical negation and conjunction. (c) Product LMDP $\mathcal{M} \times \mathcal{A}$, where red nodes represent states in the accepting MEC of $\mathcal{M} \times \mathcal{A}$. The blue transitions define the product LMC $(\mathcal{M} \times \mathcal{A})_\pi$ induced by the policy $\pi$. Note that this policy induces an LMC that has probability 1 of being trapped in the accepting MEC. Therefore, the probability of satisfying the $\omega$-regular property represented by $\mathcal{A}$ given that we start in state $s_0$ is 1.}
\label{fig:prod}
\end{center}
\end{figure}


Given an LMDP $\mathcal{M}$ and an $\omega$-regular objective $\varphi$ given as a DRA
$\mathcal{A} = (\Sigma,Q,q_0,$ $\delta, F)$, the controller synthesis problem is
to compute a policy that maximizes the probability of satisfaction of the
$\omega$-regular objective. This problem is typically reduced to solving a
product LMDP as shown in Figure~\ref{fig:prod}. Given an LMDP $\mathcal{M} = (S, \beta, A, T, R, AP, L)$ and a DRA $\mathcal{A} = (2^{AP}, Q, q_0, \delta, F)$, their \emph{product} LMDP $\mathcal{M} {\times} \mathcal{A}$ is the tuple $(S^\times, \beta^\times, A^\times, T^\times, R^\times, Q, L^\times)$, where $S^\times = S \times Q$; $\beta^\times \in D(S^\times)$ is such that for all $(s, a) \in S^\times$, we have that $\beta^\times(s, q)$ equals $\beta(s)$  if $q = \delta(q_0, L(s))$ and is $0$ otherwise;
$A^\times = A$ and  $A^\times(s, q) = A(s)$ for all $(s, q) \in S^\times$; $T^\times: S^\times {\times} A^\times \mapsto S^\times$ is such that for all $(s, q), (s', q') \in S^\times$ and $a \in A(s, q)$ we have
$T^\times((s, q), a)(s', q')$ equals $T(s, a)(s')$ if $q' = \delta(q, L(s'))$ and is $0$ otherwise;
$R^\times((s, q), a) = R(s, a)$ for all $(s, q) \in S^\times$ and $a \in  A(s, q)$; and $L^\times((s, q)) = \set{q}$ for all $(s, q) \in S^\times$ \cite{principlesOfModelChecking}.

End components and runs are defined for products just like for LMDPs. The acceptance condition for the product LMDP can be lifted from the DRA and is used to define accepting MECs (AMECs). An AMEC of a product LMDP $\mathcal{M} \times \mathcal{A}$ is a MEC such that every run of the product LMDP that eventually dwells in it is accepting. Formally, a MEC $E = (S^E, A^E)$ of $\mathcal{M} \times \mathcal{A}$ is accepting if  $S^E \cap (S \times B) = \emptyset$ and $S^E \cap (S \times G) \neq \emptyset$ for some $(B, G) \in F$.
The satisfaction of an $\omega$-regular objective $\varphi$ by an LMDP $\mathcal{M}$ can be formulated in terms of AMECs of the product $\mathcal{M} \times \mathcal{A}_\varphi$, where $\mathcal{A}_\varphi$ is a DRA accepting $\varphi$.  The maximum probability of satisfaction of $\varphi$ by $\mathcal{M}$ is the maximum probability, over all policies, that a run of the product LMDP $\mathcal{M} \times \mathcal{A}_\varphi$ eventually dwells in one of its AMECs \cite{deAlfa98,maxProbs}.
%
%
%
Once an AMEC is reached, one must simply choose actions in the AMEC infinitely often in order to ensure that all states within it are visited infinitely often. It is worth noting that there always exists a stationary and deterministic policy over the product LMDP $\mathcal{M} \times \mathcal{A}$ to maximize the probability of visiting AMECs. This policy defines the optimal finite-memory policy over the original LMDP $\mathcal{M}$ to satisfy the $\omega$-regular objective given by the DRA $\mathcal{A}$. The DRA states $q \in Q$ in $\pi: S \times Q \rightarrow A$ in the product LMDP naturally define the memory mode of the finite-memory policy in the original LMDP \cite{positionalToFinite}.

\vspace{1em}
\noindent\textbf{Steady-State Constraints.}
Let $\mathcal{M} =  (S, \beta, T, R, AP, L)$ be an LMC. A state $s'\in S$ in $\mathcal{M}$ is {\em reachable} from a state $s \in S$, denoted by $s\hookrightarrow s'$, if there exists a run $\seq{s_i, s_j, \ldots, s_k} \in S^+$ such that $s_i = s$, $s_k = s'$, and for all $0 \leq i < k$ we have that $T(s_i)(s_{i+1}) > 0$. We say that two states $s, s' \in S$ communicate if $s \hookrightarrow s'$ and $s' \hookrightarrow s$. A Markov chain is {\em irreducible} if every pair of states $s, s' \in S$ communicates. A state $s \in S$ is {\it recurrent} if for all states $s' \in S$ such that $s \hookrightarrow s'$, we have that $s' \hookrightarrow s$. A transient state is a state that is not recurrent.

A recurrent component $C \subseteq S$ of states is a nonempty set of states such that every state in $C$ communicates with every other state in $C$, and does not communicate with the states not in $C$. A {\it unichain} is an LMC that contains a single recurrent component and possibly some transient states. Otherwise, it is called a {\it multichain}. Our proposed approach finds a unichain LMC $\mathcal{M}_\pi$ in the original LMDP $\mathcal{M}$ that satisfies a given set of linear-time and steady-state constraints, though its corresponding product LMC $(\mathcal{M} \times \mathcal{A})_\pi$ in the product LMDP $\mathcal{M} \times \mathcal{A}$ may be a multichain.\par

The steady-state distribution $\text{Pr}^\infty \in D(S)$ of an LMC $\mathcal{M} = (S, \beta, T, R, AP, L)$ denotes the proportion of time spent in each state as the number of transitions within $\mathcal{M}$ approaches $\infty$. This distribution is characterized by the following system of steady-state equations:
\newcommand\dotsCompact{\hbox to 0.75em{.\hss.\hss.}}
\begin{equation}
\centering
\begin{aligned}
\hspace*{-8pt} (\text{Pr}^\infty (s_1), \dotsCompact, \text{Pr}^\infty (s_{|S|})) \cdot T & = (\text{Pr}^\infty (s_1), \dotsCompact, \text{Pr}^\infty (s_{|S|})) \\ \sum_{s \in S} \text{Pr}^\infty (s) & = 1
\end{aligned}
\label{steadyStateEquations}
\end{equation}
The unichain condition is sufficient for LMCs to yield solutions to the steady-state equations in system (\ref{steadyStateEquations}). In particular, solutions in such settings yield the unique stationary distribution corresponding to the true steady-state behavior of the agent. For multichain LMCs, however, solutions to these equations may not be unique and may not correspond to the true steady-state behavior of the agent \cite{norris1998markov}. Indeed, consider the following simple example. Let $M$ be a Markov chain defined over states $S = \{s_0, s_1, s_2\}$ such that $T(s_0)(s_1) = 0.6, T(s_0)(s_2) = 0.4, T(s_1)(s_1) = T(s_2)(s_2) = 1$. That is, state $s_0$ connects to $s_1$ and $s_2$ whereas these states self-loop with probability $1$. Solving the steady-state equations for this Markov chain yields the trivial identities $\text{Pr}^\infty(s_0) = 0, \text{Pr}^\infty(s_1) = \text{Pr}^\infty(s_1), \text{Pr}^\infty(s_2) = \text{Pr}^\infty(s_2)$, and the equation $\text{Pr}^\infty(s_1) + \text{Pr}^\infty(s_2) = 1$. Note that there are an infinite number of solutions to this equation. This elucidates the challenge of reasoning about steady-state constraints in multichain settings. We circumvent these challenges by focusing our attention on multichain product LMCs whose
BSCCs share some state of the original LMC. We show that this is a necessary and sufficient condition for the original LMC to be a unichain. Furthermore, this restricts the BSCCs of the product LMC to be identical to one another in that their transition matrices are
the same (up to row ordering). This yields a one-to-one correspondence
between the solutions to the steady-state equations in the product LMC and the
solutions to the steady-state equations in the original LMC. Since the original LMC is a unichain, this implies that these
solutions will reflect the true steady-state behavior of the agent. 


Given an LMDP $\mathcal{M} = (S, \beta, A, T, R, AP, L)$, the inverse of the labeling function $L^{-1}: 2^{AP} \rightarrow 2^S$ returns the states where a given set of atomic propositions holds. More generally, given a Boolean formula over atomic propositions $\psi \triangleq \textbf{true} \mid  p \in AP \mid \psi_1 \wedge \psi_2 \mid \neg \psi$, the function $L^{-1} (\psi) \subseteq S$ returns the set of states where $\psi$ holds. We now formalize what a steady-state specification is.

\begin{definition}[Steady-State Specification]
Given an LMC $\mathcal{M} = (S, \beta, T, $ $R, AP, L)$ and a Boolean formula $\psi$ over $AP$, a steady-state specification is a constraint of the form $l \leq \sum_{s \in L^{-1} (\psi)} \text{Pr}^\infty (s) \leq u$, where $l$ and $u$ are user-defined bounds. We let $\mathbf{SS}_{[l, u]} \psi$  denote such specifications.
\label{def:SS}
\end{definition}

\section{Related Work}
\label{sec:related}

The controller synthesis problem given $\omega$-regular objectives has been studied at length in the literature, particularly under the name of LTL controller synthesis \cite{etessami2007multi,yannakakis2008multi,forejt2011automated}. Traditionally, such problems are solved by efficiently computing the set of AMECs \cite{Chatte11} and finding a policy that reaches these and visits an accepting state therein infinitely often. The problem of deriving a control policy which satisfies constraints on the steady-state distribution of the resulting agent has been studied more recently \cite{SSC,ijcai2019-784,ijcai2020-563}. However, the literature on solving expected-reward constrained MDPs has often studied similar problems given that the expected-reward objective leverages the steady-state distribution or occupation measures, which are analogous to the steady-state distribution over state-action pairs, in order to determine expected policy values \cite{kallenberg1983linear,Puterman:1994,altman_total_cost_98,4927531}. However, the common assumption that all policies yield an irreducible Markov chain is adopted for these methods. Indeed, the stronger ergodic assumption is often made in average-reward reinforcement learning problems (\cite{sutton2018reinforcement}, Sections 10.3, 13.6).


While various extensions to LTL have introduced average \cite{averageLTL}, discounted \cite{discountedLTL}, mean-payoff \cite{LTL-lim}, and frequency \cite{frequencyLTL} modalities to the logic, to the best of the authors' knowledge, the two facets of asymptotic behavior given by the steady-state (SS) distribution and linear-time (LTL) behavior of the agent have not yet been incorporated for the deterministic controller synthesis problem. To reiterate one of the challenges in this SS+LTL controller synthesis, the choice of actions within AMECs is critical since it is the states within said AMECs that will contribute to the steady-state distribution of the Markov chain induced by the solution policy. All other states would be transient or not visited, yielding a steady-state probability measure of 0. While this challenge is not present in traditional controller synthesis problems, a restricted form of it is addressed in the problem of LTL controller synthesis subject to persistent surveillance costs \cite{svorevnova2013optimal}. The goal in these problems is to satisfy a given LTL formula, or some restricted logic fragment thereof, while minimizing the cost incurred between satisfactions of a given surveillance goal specified as the repeated observance of a goal state. 
Perhaps the work most relevant to the results established in this paper stems from \cite{SSC} and \cite{ijcai2019-784}. In \cite{SSC}, the Steady-State Control (SSC) problem is introduced. This is then generalized as Steady-State Policy Synthesis (SSPS) in \cite{ijcai2019-784}. In particular, the SSC problem entails finding a policy whose induced Markov chain satisfies a given steady-state distribution. This problem assumes that the underlying MDP is ergodic in that every policy yields irreducible Markov chains. This ensures that steady-state distributions reflect the true asymptotic behavior of the Markov chain. This is a fairly common assumption as observed recently by Altman in \cite{Altman2019_11} for average-reward or -cost problems in constrained MDPs. In \cite{ijcai2019-784}, the SSPS problem is posed as a generalization of SSC by allowing steady-state constraints to contain inequalities as well as probability intervals. The solution proposed therein does not assume ergodic MDPs and instead finds an irreducible Markov chain within an arbitrary MDP, if one exists, such that steady-state constraints are satisfied. However, that approach cannot handle transient states nor multichain MDPs. These issues were addressed recently in \cite{ijcai2020-563} and \cite{atia2021steady}, wherein a solution to the steady-state planning problem is proposed for multichain MDPs by focusing on a restricted class of policies, such as imposing that all actions be taken with some probability by the solution policy or that the long-term play is restricted to the bottom strongly connected (BSCCs) of the MDP. Indeed, the general problem of finding policies that satisfy arbitrary steady-state constraints in multichain MDPs remains open. This warrants an important distinction in our setting. Even though the product MDP over which we define our solution may be multichain, our setting is restricted in that we search for a policy that induces a product Markov chain whose BSCCs are isomorphic to one another in that their graph structures are identical. As we demonstrate, this is a necessary and sufficient condition for the original Markov chain (in the original MDP) to be a unichain, thereby ensuring that the steady-state equations admit a solution corresponding to the steady-state behavior of the agent. 

Our solution to what we call the SS+LTL controller synthesis problem unifies much of the foregoing by reasoning about both linear-time $\omega$-regular properties as well as steady-state distributions simultaneously. The proposed approach finds an optimal expected-reward control policy that is deterministic and satisfies the given steady-state (SS) and LTL specifications. We do not assume that the underlying MDP is ergodic nor communicating. Instead, our solution finds a unichain Markov chain satisfying the given specifications, if one exists. This complements the recent results in \cite{kvretinsky2021ltl}, where a stochastic history-dependent (possibly with unbounded memory) policy as in \cite{krass} is computed for the LTL-constrained steady-state policy synthesis problem. It is worth noting that, from a complexity perspective, these are fundamentally different problems due to the distinction between stochastic and deterministic policies. Indeed, finding a stochastic policy for this problem is in the complexity class \textbf{P} as demonstrated by the polynomial-time solution proposed in \cite{kvretinsky2021ltl}. On the other hand, the problem of computing a deterministic policy in this setting is an \textbf{NP-complete} problem \cite{ijcai2019-784}. Therefore, a polynomial-time solution is not likely to exist.

\section{SS+LTL Controller Synthesis}
\label{sec:ssltl}
We combine the linear-time and steady-state specifications and solve
the corresponding controller synthesis problem. 
%
%
Given an LMC $\mathcal{M}$ and a steady-state specification $\textbf{SS}_{[l, u]} (\psi)$, we say $\mathcal{M}$ satisfies $\textbf{SS}_{[l, u]} (\psi)$, denoted by $\mathcal{M} \models \textbf{ SS}_{[l, u]} (\psi)$, iff $\Sigma_{s\in L^{-1} (\psi)} \text{Pr}^\infty (s) \in [l, u]$ per Definition \ref{def:SS}. Given an LTL formula $\phi$ defined inductively over a set of atomic propositions $AP$, Boolean connectives, and temporal modalities $\textit{next}$, $\textit{until}$, $\textit{eventually}$, $\textit{always}$ ($\mathbf{X}$, $\mathbf{U}$, $\mathbf{F}$, $\mathbf{G}$), the satisfaction semantics $\mathcal{M} \models \phi$ are defined in the standard way \cite{principlesOfModelChecking}. We are interested in the combination of these LTL and SS specifications, henceforth referred to as SS+LTL specifications denoted by $\theta = (\phi_\text{LTL}, (\textbf{SS}_{[l_i, u_i]} \psi_i)_i)$. We say that $\mathcal{M}$ satisfies $\theta$, denoted by $\mathcal{M} \models \theta$, if $\mathcal{M} \models \phi_\text{LTL}$ and $\mathcal{M} \models \textbf{SS}_{[l_i, u_i]} \psi_i$ for all $i$.

\begin{definition}[Deterministic SS+LTL Controller Synthesis]
	\label{def:controllerSynthesis}
	Given an LMDP $\mathcal{M}$ and SS+LTL
	specification $\theta$, compute a finite-memory deterministic policy $\pi$, if
	one exists, such that $\mathcal{M}_\pi \models \theta$ and $\pi$ maximizes
	the expected reward among all such policies.
\end{definition}

\begin{figure}
\begin{center}
	\includegraphics[width=1.01\columnwidth]{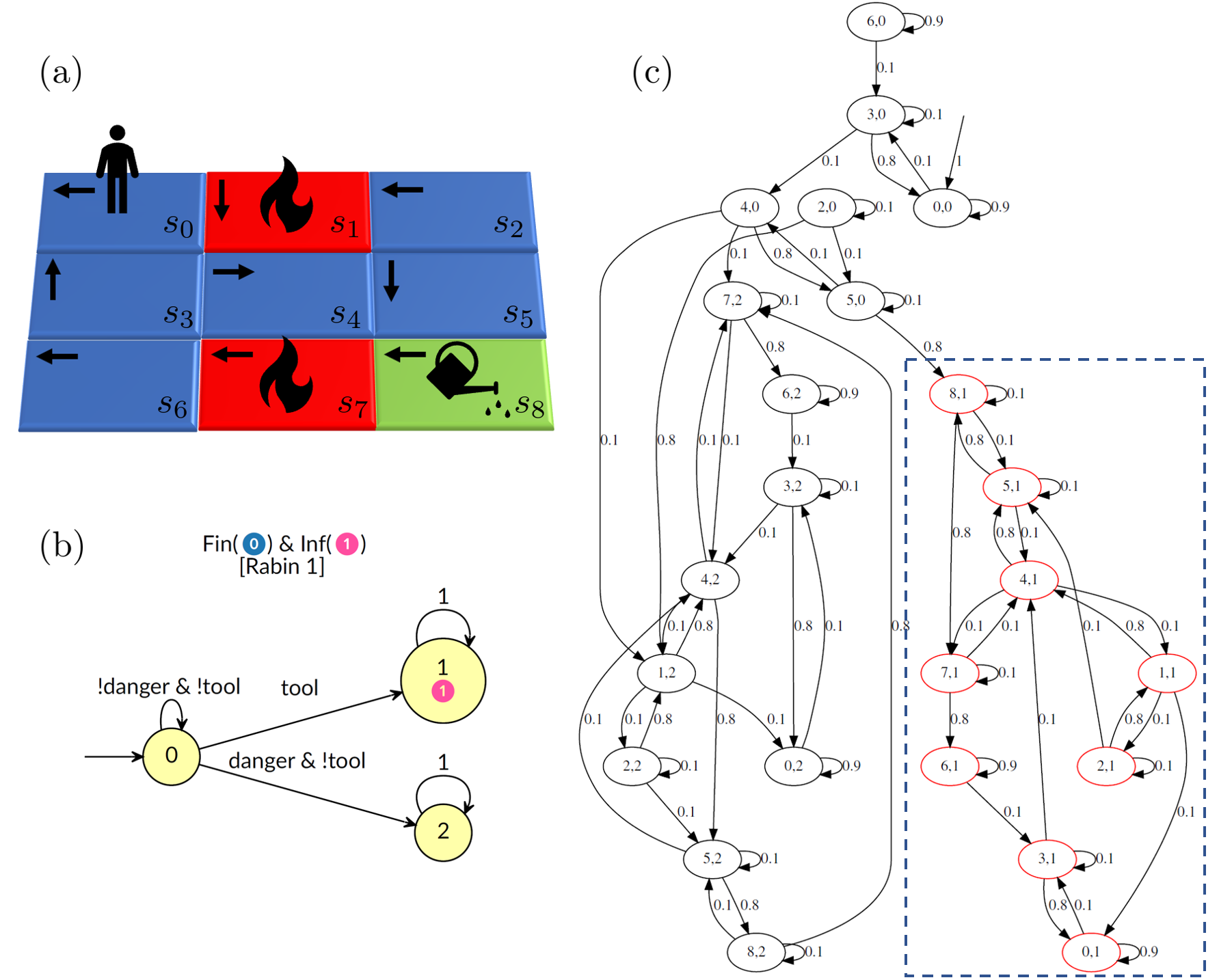}
	\caption{(a) LMDP $\mathcal{M} = (S, \beta, A, T, R = \emptyset, AP, L)$, where $S = \{s_0, \dots, s_8\}$, $\beta(s_0) = 1$, $A = \{\leftarrow, \downarrow, \rightarrow, \uparrow\}$, $AP = \{home, danger, tool\}$, and $L(s_0) = \{home\}, L(s_1) = L(s_7) = \{danger\}, L(s_8) = \{tool\}$. the agent has a chance of slipping whenever it moves, causing a transition into one of three possible states. If the agent chooses to go right (left), there is an 80$\%$ chance that it will transition to the right (left), and the chance of transitioning to either of the states above or below it is 10$\%$ each. Similarly, if the agent chooses to go up (down), it will end up in the states above (below) it with 80$\%$ chance, and in the states to the right and left of it with probability 10$\%$ each. In the corners of the map, the agent may stay in place with 90$\%$ probability by choosing to move against the boundary of the map (e.g. $T(s_0, \leftarrow)(s_0) = 0.9$). (b) Given the SS+LTL specification $\theta = ((! danger \textbf{U} tool), \textbf{SS}_{[0.75, 1]} home)$, the corresponding LTL DRA $\mathcal{A} = (Q, q_0, \Sigma, \delta, F = \{(B_i, G_i)\}_i)$ is defined, where $Q = \{q_0, q_1, q_2\}$, $\Sigma = 2^{AP}$, $F = \{(\emptyset, \{q_1\})\}$, and the transition function is given by $\delta(q_0, \emptyset) = q_0, \delta(q_0, \{tool\}) = q_1, \delta(q_0, \{danger\}) = q_2, \delta(q_1, \cdot) = q_1, \delta(q_2, \cdot) = q_2$. The symbols $!, \&$ denote logical negation and conjunction. Note that the steady-state specification in $\theta$ is not used in defining $\mathcal{A}$. (c) Product LMC $(\mathcal{M} \times \mathcal{A})_\pi$ induced by the policy $\pi$ given by the black arrows in (a) for the product LMDP $\mathcal{M} \times \mathcal{A}$, where red nodes represent states in the accepting BSCC of $(\mathcal{M} \times \mathcal{A})_\pi$. Note that this policy has non-zero probability of being trapped in the accepting BSCC. Furthermore, note that $\sum_{s \in L^{-1}(home)} \text{Pr}^\infty_\pi (s) = \text{Pr}^\infty_\pi (s_0) = 0.76$, thereby satisfying the steady-state operator $\textbf{SS}_{[0.75, 1]} home$. In this example, the product LMC is a multichain; however, note that the original LMC over states $s \in S$ as given by the dashed component (ignoring the $q \in Q$ in each $(s, q) \in S^\times$) is a unichain. Furthermore, the two BSCCs of the multichain product LMC are identical with respect to their transition matrices due to the one-to-one correspondence of paths in the original LMC and the product LMC.}
	\label{fig:ss+ltl}
\end{center}
\end{figure}

Let us fix an LMDP $ \mathcal{M} = (S, \beta, A, T, R, AP, L)$ and an SS+LTL
specification $\theta = (\phi_\text{LTL}, (\textbf{SS}_{[l_i, u_i]} \psi_i)_i)$
for the rest of the paper. Recall that the LTL formula $\phi_\text{LTL}$ can be
compiled into a DRA $\mathcal{A}$. In what follows, we work with the product LMDP
$\mathcal{M} \times \mathcal{A} = (S^\times, \beta^\times, A^\times, T^\times, R^\times, Q,
L^\times)$, sometimes referred to as $\mathcal{M}^\times$ for convenience. Our goal is to
characterize the existence of a stationary and deterministic policy $\pi: S
\times Q \rightarrow A$ over the product LMDP. This, in turn, is equivalent
to a finite-memory deterministic policy over the original LMDP. 
See Figure~\ref{fig:ss+ltl} for an example. 

\section{Integer Linear Program Characterization}
\label{sec:results}

Let us first consider an agent whose goal is to find a stationary stochastic policy $\pi: S \times Q \rightarrow D(A)$ to maximize the expected reward in a product LMDP $\mathcal{M} \times \mathcal{A}$. If $\mathcal{M} \times \mathcal{A}$ is a unichain LMDP, the program in system (\ref{program:original}) suffices to compute the optimal policy such that the solution yields the identity $x_{sqa} = \pi(a | s, q) \text{Pr}^{\infty} (s, q) = \pi(a | s, q) \sum_a x_{sqa}$ for $s \in S$, $q \in Q$, and $a \in A$ from which a stochastic policy can then be derived, where $x_{sqa}$ denotes the occupation measure of taking action $a$ in state $(s, q) \in S^\times$ \cite{occupationMeasure}.
\begin{gather}
\begin{aligned}
& \max \sum_{(s, q) \in S^\times} \sum_{a \in A(s)} x_{sqa} \! \sum_{s' \in S} T(s, a)(s') R(s, a, s') \text{ subject to} \\  
& (i) \sum_{(s, q) \in S^\times} \sum_{a \in A(s)} x_{sqa} T^\times ((s, q), a )(s', q') = \!\!\! \sum_{a \in A(s')} x_{s'q'a} \\ & \hspace*{154pt} \forall (s', q') \in S^\times \\ 
& (ii) \sum_{(s, q) \in S^\times} \sum_{a \in A(s)} x_{sqa} = 1 \\
\end{aligned}
\label{program:original}
\end{gather}

Now, consider the more general case where the given product LMDP $\mathcal{M} \times \mathcal{A}$ may be multichain. Two key problems arise. First, the policy $\pi$ derived from the solution to (\ref{program:original}) may not yield a unichain original LMC $\mathcal{M}_\pi$ (i.e., one with a single BSCC and possibly some transient states). Second, we note the challenges of deriving the correct steady-state distributions for an agent using linear programming in the multichain setting. In particular, in his seminal work \cite{kallenberg1983linear}, Kallenberg demonstrated that there is not a one-to-one correspondence between the steady-state distribution derived from linear programming solutions to expected-reward MDPs and the true steady-state distribution of the agent enacting the resulting policy when the Markov chain is multichain (i.e. contains multiple BSCCs, and possibly some transient states). On the other hand, unichains yield a one-to-one correspondence between the solution of the steady-state equations and the true steady-state behavior of the agent \cite{Puterman:1994}. Furthermore, the solution to these equations is unique in said setting.  We thus focus on deriving an optimal solution policy $\pi: S \times Q \rightarrow A$ in a (potentially) multichain product LMDP $\mathcal{M} \times \mathcal{A}$ such that the induced original LMC $\mathcal{M}_\pi$ is a unichain and satisfies a given SS+LTL formula. The interplay with the product LMC $(\mathcal{M} \times \mathcal{A})_\pi$ introduces some challenges in deriving such a policy. In particular, it may be the case that the product LMC $(\mathcal{M} \times \mathcal{A})_\pi$ induced by the solution policy $\pi$ is multichain and its corresponding original LMC $\mathcal{M}_\pi$ is unichain. We present a novel solution which accounts for such settings by ensuring that all BSCCs in the product multichain $(\mathcal{M} \times \mathcal{A})_\pi$ share some state of the original LMC $\mathcal{M}_\pi$. This establishes that the original LMC $\mathcal{M}_\pi$ is a unichain. We further prove that the steady-state probabilities derived over such product LMCs yield a one-to-one correspondence with the true steady-state behavior of the agent in the original LMC. 



First, let us consider the simpler case where the product LMC is a unichain. Note that the single BSCC may contain the same state $s \in S$ multiple times as $S^\times \ni (s, q), (s, q'), \dots$ We will show that the partition $[s] = \{(s, q)\}_q$ naturally defined over the states of the product LMC to yield the states of the original LMC is such that $\text{Pr}^\infty (s) = \sum_{(s, q) \in [s]} \text{Pr}^\infty (s, q)$. That is, we can compute the steady-state probabilities over the product LMC and use these to derive those in the original LMC over which the SS+LTL specification is defined. This is enabled by the lumpability of the product LMC, defined below. 


\begin{definition}[Lumpability \cite{lumpability}, Def. 1]
Given an irreducible Markov chain $\mathcal{M} = (S, T)$ and a partition $\bigcup_{k=1}^K S_k$ $(S_k \subset S, S_i \cap S_j = \emptyset)$ of $S$, then $\bigcup_{k} S_k$ is called ordinarily lumpable if and only if $(\mathbf{e}_\alpha -\mathbf{e}_\beta) T V = \mathbf{0}$ for all $s_\alpha, s_\beta \in S_k, k \leq K$, where $\mathbf{e}$ is the standard basis vector and $V$ is defined so that $v_{ik} = 1$ if $s_i \in S_k$ and $v_{ik} = 0$ otherwise. The vector $\mathbf{e}_k$ is the all-zeroes vector with a value of 1 only for the $k^\text{th}$ entry. 
\end{definition}

A partition over an LMC naturally defines another LMC, known as the aggregated LMC, where each state of the latter corresponds to one of the partition sets of the former. As we will show in Corollary 1, the original LMC is the aggregated LMC resulting from a lumpable partition of the product LMC. 

\begin{definition}[Aggregated Markov Chain]
    Given a product LMC $\mathcal{M}^\times = (S^\times = S \times Q, \beta^\times,$ $ T^\times, R^\times, AP, L^\times)$ and a partition $\bigcup_{s \in S} [s]$ such that $[s] = \{(s, q) | (s, q) \in S^\times\}$, the aggregated LMC is given by $\mathcal{M}^* = (S^*, \beta^*, T^*, R^*, AP, L^*)$, where $S^* = \{s | [s] \in \bigcup_{s \in S} [s]\}, \beta^* (s) = \sum_{(s, q) \in [s]} \beta^\times (s, q), T^* (s)(s') = T^\times (s, \cdot)(s', \cdot)$, $R^* (s, s') = \\ R^\times ((s, \cdot), (s', \cdot))$, and $L^* (s) = L^\times ((s, \cdot))$.
    \label{def:quotient}
\end{definition}


\begin{lemma}
Given an arbitrary BSCC $(\hat{S}, \hat{T})$ of a product LMC $\mathcal{M}^\times = (S^\times = S \times Q, T^\times)$, the partition $\bigcup_{s \in S} [s]$ given by equivalence classes $[s] = \{(s, q) | (s, q) \in \hat{S}\}$ is ordinarily lumpable. The proof is in Appendix~\ref{sec:ordinarily-lumpable}. 
\label{ordinarily-lumpable}
\end{lemma}

\noindent We adapt a theorem from \cite{lumpability} and modify it for our product LMC setting below.


\begin{theorem}[\cite{lumpabilityOriginal}, \cite{lumpability}, Theorem 4]
Given an irreducible product LMC $\mathcal{M}^\times = (S^\times = S \times Q, T^\times)$ and an ordinarily lumpable partition $\bigcup_{s \in S} [s]$, where $[s] = \{(s, q) |$ $ (s, q) \in S^\times\}$, the steady-state distribution of the aggregated LMC $\mathcal{M} = (S, T)$ satisfies $\text{Pr}^\infty (s) = \sum_{(s, q) \in [s]} \text{Pr}^\infty (s, q)$ for every $s \in S$. Furthermore, the transition function of the aggregated LMC is given by $T(s)(s') = \mathbf{e}_i T([s])([s']) \mathbf{e}^T$, where $i$ is an arbitrary index in the set $\{i | (s_i, \cdot) \in [s]\}$. The proof is in Appendix B. 
\label{theorem:aggregateLMC}
\end{theorem}

\begin{corollary}
Given an irreducible product LMC $\mathcal{M}^\times = (S^\times, T^\times)$, the original LMC $\mathcal{M} = (S, T)$ is the aggregated LMC resulting from the ordinarily lumpable partition $\bigcup_{s \in S} [s]$, where $[s] = \{(s, q) | (s, q) \in S^\times\}$. The proof is in Appendix B. 
\label{corollary:aggregateIsOriginal}
\end{corollary}
Lemma \ref{ordinarily-lumpable}, Theorem \ref{theorem:aggregateLMC}, and Corollary \ref{corollary:aggregateIsOriginal} establish the one-to-one correspondence between the steady-state probability derived for an irreducible product LMC and the steady-state distribution for the original LMC. Note that this result also holds for unichains since the steady-state probability measure of transient states therein would be zero. Now, let us consider the case where the product LMC is a multichain. We establish sufficient conditions for yielding the same one-to-one correspondence of steady-state distributions. Furthermore, we establish necessary and sufficient conditions for the multichain product LMC $(\mathcal{M} \times \mathcal{A})_\pi$ to yield a unichain original LMC $\mathcal{M}_\pi$.


\begin{lemma}
Let $\mathcal{M}^\times = (S^\times, T^\times)$ denote a multichain product LMC and let $(S^k)_k, S^k \subset S^\times$ denote its BSCCs. Then its corresponding original LMC $\mathcal{M} = (S, T)$ is a unichain iff some state $(s, \cdot) \in S^\times$ shows up in every BSCC $S^k$ of $\mathcal{M}^\times$. That is, for some $s \in S$ and all $k$, there exists $q \in Q$ such that $(s, q) \in S^k$.
\label{lemma:multiToUni}
\end{lemma}

\begin{proof}
This follows from the one-to-one correspondence between paths in $\mathcal{M}$ and paths in $\mathcal{M}^\times$. Furthermore, the single BSCC $S' \subseteq S$ of $\mathcal{M}$ is given by $S' = \{s | (s, q) \in \bigcup_k S^k\}$.
\end{proof}

\begin{lemma}
    Given a multichain product LMC $\mathcal{M}^\times = (S^\times, T^\times)$ with $m$ identical BSCCs given by transition probability matrices $T_1 = T_2 = \dots = T_m$, and an irreducible LMC $\mathcal{M}' = (S', T')$, where $S'$ contains exactly the states in the first BSCC and $T' = T_1$ (w.l.o.g.), the steady-state probability of an arbitrary state $(s, q) \in S'$ is equivalent to the sum of steady-state probabilities of all states isomorphic to it in the BSCCs of $\mathcal{M}^\times$. The proof is in Appendix~\ref{sec:key-lemma}. 
    \label{lemma:key}
\end{lemma}

\begin{figure}[h]
    \centering
	\includegraphics[width=8cm]{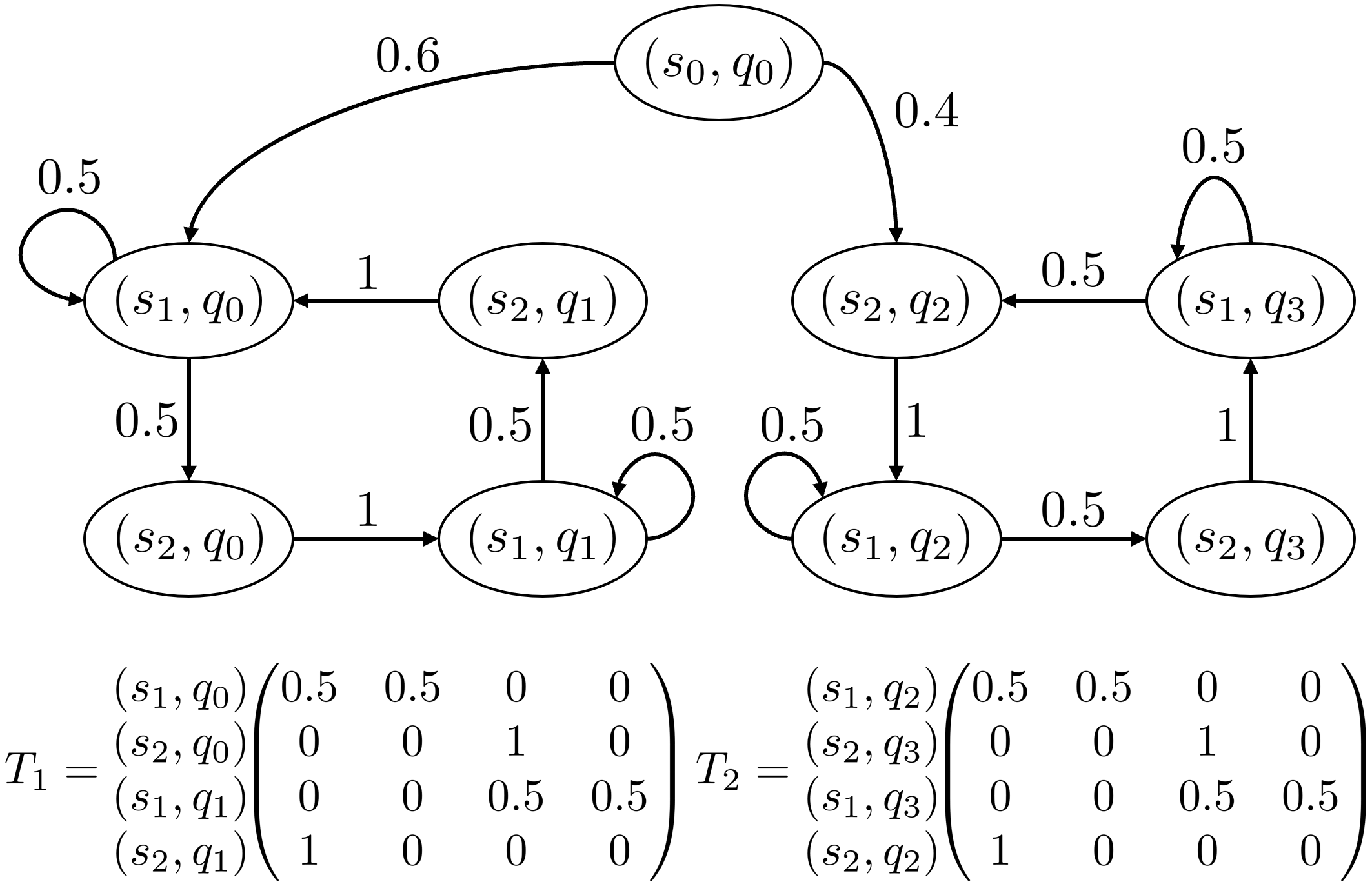}
  \begin{minipage}[c]{0.4\textwidth}
	\caption{Product LMC with isomorphic sets given by $\langle s_1 \rangle = \{(s_1, q_0), (s_1, q_2)\}$, $\langle s'_1 \rangle = \{(s_1, q_1), (s_1, q_3)\}$, $\langle s_2 \rangle = \{(s_2, q_0), (s_2, q_3)\}$, $\langle s'_2 \rangle = \{(s_2, q_1), (s_2, q_2)\}$.}
	\label{fig:isomorphic}
	\includegraphics[width=7cm]{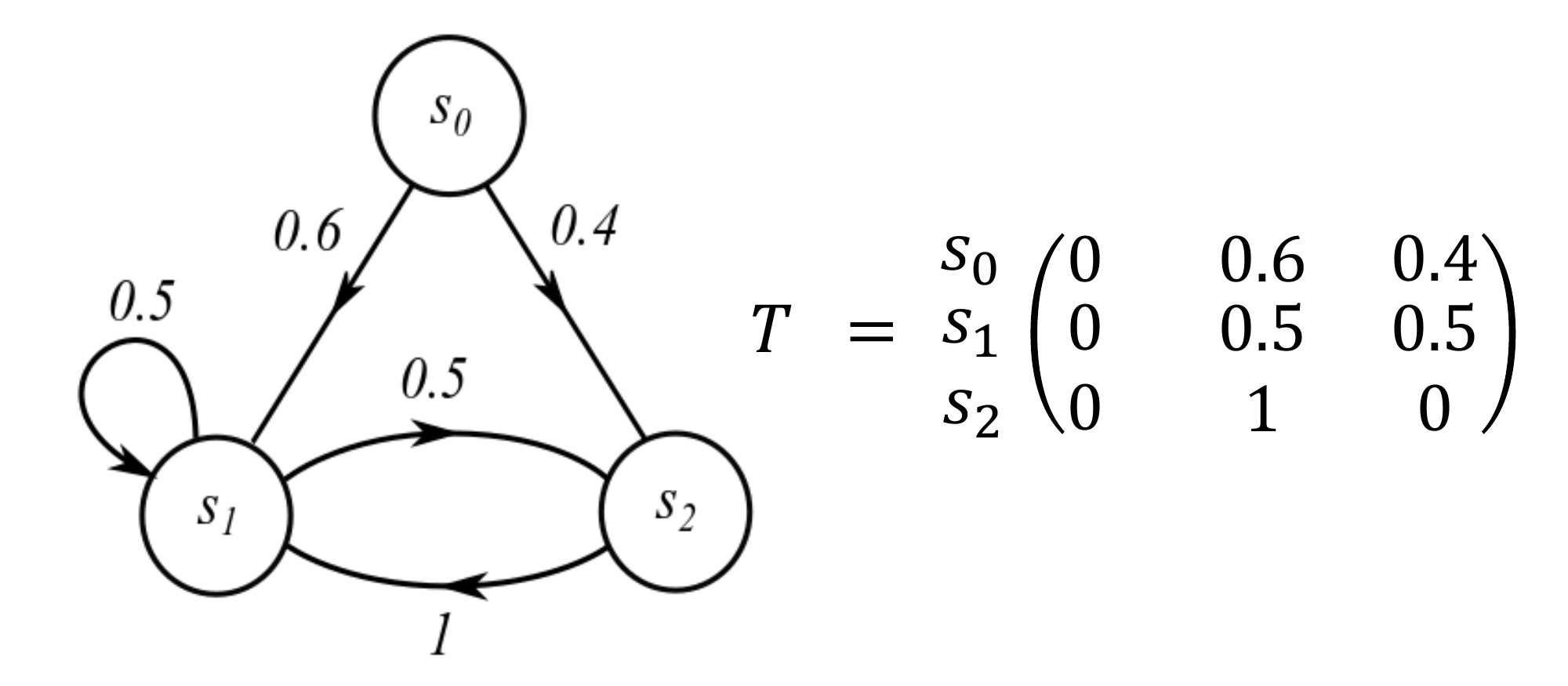}
	\caption{The unichain original LMC.}
	\label{fig:isomorphicOriginalLMC}
  \end{minipage}
\end{figure}

To illustrate Theorem \ref{theorem:aggregateLMC} and Lemmas \ref{lemma:multiToUni} and \ref{lemma:key}, consider the multichain product LMC in Figure \ref{fig:isomorphic}, where $T_1$ and $T_2$ denote the transition probability matrices for the two BSCCs. We will show that, because these two BSCCs are identical in terms of their transition matrices (rows may need to be reordered to reflect this), we have $\text{Pr}^\infty (s) = \sum_{(s, q) \in [s]} \text{Pr}^\infty (s, q)$ for states $s$ in the original LMC shown in Figure \ref{fig:isomorphicOriginalLMC}. The solution to the steady-state equations for this product LMC yields $\text{Pr}^\infty (s_0, q_0) = 0, \text{Pr}^\infty (s_1, \cdot) = 1 / 6, \text{Pr}^\infty (s_2, \cdot) = 1 / 12$. The order of states in $T_2$ differs from that of $T_1$ in order to reflect that BSCCs can be identical up to row ordering. Note that the equivalence classes can be defined in terms of the isomorphic sets as $[s_1] = \langle s_1 \rangle \bigcup \langle s'_1 \rangle$ and $[s_2] = \langle s_2 \rangle \bigcup \langle s'_2 \rangle$. 

\setlength{\textfloatsep}{8pt}
\setlength{\intextsep}{8pt}

Now, consider the original LMC shown in Figure \ref{fig:isomorphicOriginalLMC} 
corresponding to this product LMC.
Solving the steady-state equations (\ref{steadyStateEquations}) for the original LMC yields $\text{Pr}^\infty (s_0) = 0, \text{Pr}^\infty (s_1) = 2 / 3, \text{Pr}^\infty (s_2) = 1 / 3$. Note that $\text{Pr}^\infty (s_1) = \sum_{(s, q) \in [s_1]} \text{Pr}^\infty (s, q)$ and $\text{Pr}^\infty (s_2) = \\ \sum_{(s, q) \in [s_2]} \text{Pr}^\infty (s, q)$ in accordance with Theorem \ref{theorem:aggregateLMC} by leveraging the fact that $\text{Pr}^\infty (s, q) = \sum_{(s', q') \in \langle s \rangle} \text{Pr}^\infty (s', q')$ per Lemma \ref{lemma:key}.



    
Theorem \ref{theorem:aggregateLMC} and Lemma \ref{lemma:key} establish necessary and sufficient conditions for a multichain product LMC to yield a unichain original LMC in the original LMDP
such that there is a one-to-one correspondence between the sum of steady-state probabilities $\sum_q \text{Pr}^\infty (s, q)$ in the former and the
steady-state distribution $\text{Pr}^\infty (s)$ in the latter. Indeed, note that 
\begin{equation}
\nonumber
\text{Pr}^\infty (s) = \sum_{(s, q) \in [s]} \text{Pr}^\infty (s, q) = \sum_{s \in S} \sum_{(s', q') \in \langle s \rangle} \text{Pr}^\infty (s', q')
\end{equation}
holds when the product LMC is unichain or multichain given that the original LMC is unichain. This is the case when all BSCCs in the product LMC are identical as mentioned in Lemma \ref{lemma:key}, which is the case when all BSCCs in the product LMC share some state in $S$ per Lemma \ref{lemma:multiToUni}.

We can now add constraints to program (\ref{program:original}) to ensure that the solution policy $\pi$ is deterministic and yields a unichain original LMC $\mathcal{M}_\pi$ in the original LMDP $\mathcal{M}$ even though the product LMC $(\mathcal{M} \times \mathcal{A})_\pi$ induced in the product LMDP $\mathcal{M} \times \mathcal{A}$ may be multichain. We begin with constraints to enforce a deterministic policy. Constraint $(iii)$ ensures that a positive occupation measure implies that the action corresponding to it is selected as part of the solution policy and $(iv)$ enforces a valid probability distribution, where $\pi(a | s, q) \in \{0, 1\}$.
\begin{gather}
\nonumber
\begin{aligned}
(iii) \text{ } & x_{sqa} \leq \pi(a | s, q) & \forall (s, q) \in S^\times, a \in A \\ 
(iv) \text{ } & \sum_{a \in A(s)} \pi \left(a | s, q \right) = 1 & \forall (s, q) \in S^\times 
\end{aligned}
\end{gather}

\begin{lemma}
	Let $(x, \pi)$ denote a feasible solution to constraints $(i)$ through $(iv)$ and assume that the product LMC $(\mathcal{M} \times \mathcal{A})_\pi$ induced by $\pi$ is such that all BSCCs share some state $s \in S$. Then $x_{sqa} = \pi (a | s, q) \text{Pr}^\infty_\pi (s, q) = \pi (a | s, q) \sum_a x_{sqa}$ for all recurrent states $(s, q) \in S^\times$. The proof is in Appendix~\ref{sec:key2}. 
	\label{lemma:key2}
\end{lemma}

We can now add additional constraints that utilize the policy $\pi$ in constraints $(iii)$ and $(iv)$ in order to establish that some accepting state within an AMEC is reached by this policy and visited infinitely often. This would, in turn, satisfy the LTL specification $\phi_\text{LTL}$ of the given SS+LTL specification $\theta$. For simplicity, we assume an initial state $s_0$ in the underlying LMDP $\mathcal{M}$ (i.e. $\beta(s_0) = 1$). In order to ensure that there is a path from the initial state $(s_0, \delta(q_0, L(s_0))) \in S^\times$ in the product LMDP $\mathcal{M}^\times$ to some recurrent component in the union of AMECs which contains nodes in $\bigcup_i G_i$ (i.e. nodes that are part of the DRA acceptance pairs), we will use flow transfer constraints. This notion of flow reflects the probability of transitioning between states given a policy. Constraint $(v)$ sets the flow capacities, where $f_{sqs'q'}$ denotes flow from $(s, q) \in S^\times$ to $(s', q') \in S^\times$. Constraint $(vi)$ ensures that, for every state (except the starting state), if there is incoming flow, then it is strictly greater than the outgoing flow. This is handled by the product of some small constant $\epsilon$ and an indicator variable $\mathcal{I}_{sq} \in \{0, 1\}$ denoting whether flow is being transferred from state $(s, q)$ to some other state. If there is no incoming flow, then there is no outgoing flow and $\mathcal{I}_{sq}$ must necessarily be zero. Constraint $(vii)$ ensures that, if there is incoming flow into a state $(s, q) \in S^\times$, then $\mathcal{I}_{sq} = 1$. Constraint $(viii)$ ensures that, whenever there is incoming flow, there must also be some arbitrary amount of outgoing flow. The choice of denominator 2 here is arbitrary. 
\begin{gather}
\nonumber
\begin{aligned}
(v) \text{ } & f_{sqs'q'} \leq \sum_{a \in A(s)} T((s, q), a)(s', q') \pi(a | s, q) \\ & \hspace*{136pt} \forall ((s, q), (s', q')) \in T^G \\
(vi) \text{ } & \sum_{((s', q'), (s, q)) \in T^G} \!\!\!\!\! f_{s'q'sq} \geq \!\!\!\!\! \sum_{((s, q), (s', q')) \in T^G} \!\!\!\!\! f_{sqs'q'} + \epsilon \mathcal{I}_{sq} \\ & \hspace*{96pt} \forall (s, q) \in S^\times \setminus \{(s_0, \delta(q_0, L(s_0)))\} \\ 
(vii) \text{ } & \sum_{((s', q'), (s, q)) \in T^G} \!\!\!\!\! f_{s'q'sq} \leq \mathcal{I}_{sq} \hspace*{74pt} \forall (s, q) \in S^\times \\ 
(viii) \text{ } & \sum_{((s, q), (s', q')) \in T^G} \!\!\!\!\! f_{sqs'q'} \geq \!\!\!\!\! \sum_{((s', q'), (s, q)) \in T^G} \!\!\!\!\! f_{s'q'sq} / 2 \hspace*{12pt} \forall (s, q) \in S^\times  
\end{aligned}
\end{gather}
Constraint $(ix)$ ensures that the steady-state probability of states with no incoming flow (as determined by $\mathcal{I}_{sq}$ in constraint $(vi)$) is 0. This makes it so that unreachable BSCCs in the product LMC $(\mathcal{M} \times \mathcal{A})_\pi$ do not contribute to the steady-state distribution. Constraint $(x)$ encodes the steady-state specifications given by operators \textbf{SS} in $\theta$ and constraint $(xi)$ ensures that some state in the acceptance sets $\bigcup_i G_i$ of $\mathcal{A}$ is visited infinitely often to satisfy the LTL specification. 
\begin{gather}
\nonumber
\begin{aligned}
(ix) \text{ } & \sum_{a \in A(s)} x_{sqa} \leq \mathcal{I}_{sq} & \forall (s, q) \in S^\times \\ 
(x) \text{ } & l \leq \sum_{s \in L^{-1}(\psi)} \sum_{q \in Q} \sum_{a \in A(s)} x_{sqa} \leq u & \forall \textbf{SS}_{[l, u]} \psi \in \theta \\ 
(xi) \text{ } & \sum_{s \in S} \sum_{q \in \bigcup_i G_i} \sum_{a \in A(s)} x_{sqa} > 0 \\
\end{aligned}
\end{gather}


Recall that constraints $(i)$ and $(ii)$ yield the correct steady-state distribution if there is a single BSCC (per the unichain condition) or if all BSCCs in the product LMC $(\mathcal{M} \times \mathcal{A})_\pi$ are identical and the induced original LMC $\mathcal{M}_\pi$ is a unichain (per Lemma \ref{lemma:key}). 
This is the case when all BSCCs of $(\mathcal{M} \times \mathcal{A})_\pi$ share some state in $s \in S$ per Lemma \ref{lemma:multiToUni}. 
We must therefore ensure that, in the product LMC, some state in $S$ is shared (This is trivially true if there is only one BSCC).
The one-to-one correspondence of paths between the original LMC and the product LMC will then guarantee that the former is unichain even if the latter is multichain. 
To accomplish this, we define three indicator variables $\mathcal{I}_s, \mathcal{I}^k, \mathcal{I}^k_s \in \{0, 1\}$ whose value is 1 iff $(s, \cdot)$ shows up in some BSCC of the product LMC, the $k^\text{th}$ AMEC of the product LMDP has some state with positive steady-state probability (meaning that the AMEC, or a subset of it, will show up as a BSCC in the product LMC), and $(s, \cdot)$ has positive steady-state probability in the $k^\text{th}$ AMEC, respectively. 

Let $AMEC$ denote the set of all AMECs in the product LMDP $\mathcal{M} \times \mathcal{A}$ and let $AMEC_k$ denote the $k^\text{th}$ AMEC. Constraint $(xii)$ ensures that $\mathcal{I}^k$ is 1 if some state in the $k^\text{th}$ AMEC has positive steady-state probability. Constraints $(xiii)$ and $(xiv)$ ensure that, for a given state $s \in S$ and the $k^\text{th}$ AMEC, some indicator variable $\mathcal{I}_{sq}$ is 1 for $(s, q)$ in the $k^\text{th}$ AMEC if and only if $\mathcal{I}^k_s = 1$. Constraint $(xv)$ ensures that, if $\mathcal{I}_s$ is 1, then $(s, \cdot)$ shows up in every BSCC of the product LMC $(\mathcal{M} \times \mathcal{A})_\pi$, thereby enforcing that all BSCCs in $(\mathcal{M} \times \mathcal{A})_\pi$ are identical. Per Lemma \ref{lemma:multiToUni}, this ensures that the original LMC $\mathcal{M}_\pi$ is a unichain. Note that the sum $\sum_k (\mathcal{I}^k_s - \mathcal{I}^k)$ is always non-positive and dividing by the number of AMECs bounds this result to be within $[-1, 0]$. Finally, constraint $(xvi)$ ensures that some such shared state exists across all BSCCs of the product LMC.
\begin{gather}
\nonumber
\begin{aligned}
(xii) \text{ } & \sum_{(s, q) \in AMEC_k} \sum_a x_{sqa} \leq \mathcal{I}^k & \forall 1 \leq k \leq |AMEC| \\ 
(xiii) \text{ } & \mathcal{I}^k_s \leq \sum_{(s, q) \in AMEC_k} \mathcal{I}_{sq} & \forall s \in S, 1 \leq k \leq |AMEC| \\ 
(xiv) \text{ } & \sum_{(s, q) \in AMEC_k} \frac{\mathcal{I}_{sq}}{|Q|} \leq \mathcal{I}^k_s & \forall s \in S, 1 \leq k \leq |AMEC| \\ 
(xv) \text{ } & \mathcal{I}_s - 1 \leq \frac{\sum_k \left( \mathcal{I}^k_s - \mathcal{I}^k \right)}{|AMEC|} & \forall s \in S \\
(xvi) \text{ } & \sum_s \mathcal{I}_s \geq 1 \\ 
\end{aligned}
\end{gather}

\noindent The program is summarized below. 
\begin{equation}
\begin{aligned}
& \max \!\!\! \sum_{(s, q) \in S^\times} \sum_{a \in A(s)} \!\! x_{sqa} \sum_{s' \in S} T(s, a)(s') R(s, a, s') \text{ s.t. } (i) - (xvi) \\ 
& x_{sqa}, f_{sqs'q'} \in [0, 1], \hspace*{24pt} \forall ((s, q), a, (s', q')) \in S^\times \!\! \times \! A \! \times \! S^\times \\ 
& \pi(a | s, q), \mathcal{I}_{sq}, \mathcal{I}_s, \mathcal{I}^k, \mathcal{I}^k_s \in \{0, 1\}, \hspace*{24pt} \\ & \hspace*{84pt} \forall ((s, q), a) \in S^\times \!\! \times \! A, 1 \leq k \leq |AMEC| \\ 
\end{aligned}
\label{equation:finalProgramss}
\end{equation}


\begin{theorem}
\label{thm:main}
    Given an LMDP $\mathcal{M} = (S, \beta, A, T, R, AP, L)$ and an SS+LTL objective $\theta = (\phi_\text{LTL}, (\textbf{SS}_{[l_i, u_i]} \psi_i)_i)$, let $(x, f, \pi, \mathcal{I})$ denote an assignment to the variables in program (\ref{equation:finalProgramss}). Then  $(x, f, \pi, \mathcal{I})$ is a feasible solution if and only if $\mathcal{M}_\pi = (S_\pi, \beta, T_\pi, AP, L)$ satisfies $\theta$ and is a unichain. The proof is in Appendix~\ref{sec:theorem-main}. 
\end{theorem}

	
\section{Experimental Results}
\label{sec:experiments}	

	Simulations of program (\ref{equation:finalProgramss}) were performed using
	CPLEX version $12.8$ \cite{iILO06a} on a machine with a $3.6$ GHz Intel Core
	i7-6850K processor and $128$ GB of RAM. We generated random $4 \times 4, 8 \times 8$, and $16 \times 16$ gridworld environments given by the LMDP $\mathcal{M} = (S, \beta, T, R, AP, L)$ subject to various SS+LTL specifications $\theta$ and with the top-left corner of the grid as the inital state. There are four actions corresponding to the four cardinal directions and a deterministic transition function $T(s, a)(s') \in \{0, 1\}$ defined in the obvious manner. Each state-action pair observes a uniformly distributed random reward in $\{0, 1\}$. See the figure in Appendix \ref{appendix:figure} for an example. It is worth noting that the solutions illustrated in Figure \ref{fig:prod} and Figure \ref{fig:ss+ltl} were also generated using program (\ref{equation:finalProgramss}).

	
	In the following experiments, the set of atomic propositions is given by $AP = \{a, b, c, d\}$, with each atomic proposition allocated to one-fourth of the states chosen at random. See Table \ref{CPLEXResults} for results. These results demonstrate that the proposed program (\ref{equation:finalProgramss}) can scale to state spaces of moderate size on the order of a few minutes.

\begin{gather}
	\nonumber
	\begin{aligned}
	    \theta_1 & = ((\mathbf{G} \neg b) \wedge (\mathbf{G} \mathbf{F} a), \textbf{SS}_{[0.01, 0.5]} d) \\ 
	    \theta_2 & = ((\mathbf{G} \mathbf{F} a) \vee (\mathbf{F} \mathbf{G} b), \textbf{SS}_{[0.01, 0.5]} d) \\ 
	    \theta_3 & = ((\mathbf{F} \mathbf{G} a) \mathbf{U} (b \vee \mathbf{X} (b \vee \mathbf{X} (b \vee \mathbf{X} b))), \textbf{SS}_{[0.01, 0.5]} d) \\ 
	    \theta_4 & = ((\mathbf{F} a) \mathbf{U} b, \textbf{SS}_{[0.01, 0.5]} d) \\ 
	    \theta_5 & = ((\mathbf{F} a) \wedge \mathbf{F} (a \mathbf{U} b), \textbf{SS}_{[0.01, 0.5]} d) \\ 
	   \theta_6 & = (\mathbf{F} (a \wedge \mathbf{X} (a \wedge \mathbf{X} a)), \textbf{SS}_{[0.01, 0.5]} d) \\ 
	   \theta_7 & = ((\mathbf{F} a \wedge \mathbf{F} b) \wedge ((\mathbf{F} a \wedge \mathbf{F} b) \mathbf{U} (c \vee \mathbf{X} a)), \textbf{SS}_{[0.01, 0.5]} d) \\ 
	   \theta_8 & = (\mathbf{F} a \wedge \mathbf{F} b \wedge \mathbf{F} c, \textbf{SS}_{[0.01, 0.5]} d) \\ 
	\end{aligned}
\end{gather}
	
\begin{table}[h] 
		\centering
		\begin{tabular}{p{1cm}p{1.5cm}p{1.75cm}p{2.5cm}}
			\toprule
			$\theta$ & $4 \times 4$ & $8 \times 8$ & $16 \times 16$ \\
			\midrule
			$\theta_1$ & 0.42 (0.43) & 13.09 (83.86) & 35.21 (70.09) \\
			$\theta_2$ & 0.28 (0.72) & 0.15 (0.06) & 1.42 (0.74) \\
			$\theta_3$ & 1.13 (3.27) & 0.72 (0.40) & 52.74 (59.42) \\
			$\theta_4$ & 0.58 (2.04) & 1.19 (0.53) & 78.29 (53.94) \\
			$\theta_5$ & 0.64 (1.93) & 1.56 (0.70) & 125.42 (93.79) \\
			$\theta_6$ & 0.25 (0.62) & 1.03 (0.43) & 155.60 (130.84) \\
			$\theta_7$ & 1.50 (5.08) & 4.95 (2.77) & 195.87 (145.07) \\
			$\theta_8$ & 2.28 (6.41) &  9.50 (6.37) & 338.88 (205.29) \\
			\bottomrule
		\end{tabular}
	    \caption{Average runtimes and standard deviations for 100 random instances of program (\ref{equation:finalProgramss}) using CPLEX version $12.8$ for the listed SS+LTL specifications $\theta_1, \dots, \theta_8$ and for grids of sizes $4 \times 4$, $8 \times 8$, and $16 \times 16$.} 
	\label{CPLEXResults}
\end{table}

\section{Conclusion}

In this paper, we proposed and solved the deterministic controller synthesis problem for labeled Markov Decision Processes (LMDPs) subject to specifications on both the linear-time and visitation frequency behaviors of an agent. The proposed approach uses a novel integer programming formulation to find a policy that induces a unichain labeled Markov chain (LMC). The program reasons about the product LMDP computed from the original LMDP and the deterministic Rabin automaton (DRA) representation of the linear-time property. Though the product LMC induced by the solution policy may be a multichain, we established necessary and sufficient conditions for the one-to-one correspondence between the visitation frequencies derived from the product LMC and the true steady-state behavior of the agent captured by the unichain original LMC. The foregoing is a step toward infinite-horizon formal synthesis of control policies in general decision processes. For future work, we will explore how similar correct-by-construction policies can be computed such that guarantees of behavior hold for general multichain LMCs induced by said policies in the original LMDP. 



\begin{acks}
This research was supported in part by the Air Force Research Laboratory through the Information Directorate’s Information Institute$^\text{®}$ Contract Number FA8750-20-3-1003 and FA8750-20-3-1004, the Air Force Office of Scientific Research through Award 20RICOR012, and the National Science Foundation through CAREER Award CCF-1552497 and Award CCF-2106339.
\end{acks}

\bibliographystyle{unsrt}
\bibliography{refs} 


\appendix 

\clearpage
\newpage



\section*{Appendix: Proofs}

\section{Proof of Lemma~\ref{ordinarily-lumpable}}
\label{sec:ordinarily-lumpable}
By definition, the BSCC $(\hat{S}, \hat{T})$ is irreducible. We must show that, for all $s \in S$ and all $(s, q_\alpha), (s, q_\beta) \in [s]$, we have $(\mathbf{e}_\alpha - \mathbf{e}_\beta) \hat{T} V = 0$. Let $V \in \{0,
1\}^{(|S||Q|) \times |S|}$ such that $v_{(ij)k} = 1$ if $(s_i, q_j) \in
[s_k]$ and $v_{(ij)k} = 0$ otherwise. It follows that:
\begin{equation}
\begin{aligned}
    (\mathbf{e}_\alpha - \mathbf{e}_\beta) \hat{T} V = \left( \hat{T}(s, q_\alpha)(s_i, q_j) -  \hat{T}(s, q_\beta)(s_i, q_j) \right)_{\substack{i \leq |S| \\ j \leq |Q|}} V\\ 
    = \left( \sum_{i \leq |S|, j \leq |Q|} v_{(ij)k} [\hat{T}(s, q_\alpha)(s_i, q_j) -  \hat{T}(s, q_\beta)(s_i, q_j)] \right)_{k \leq |S|}
\end{aligned}
\end{equation}

\noindent Looking at the $k^\text{th}$ entry, we have:

\begin{equation}
    \sum_{(s_i, q_j) \in [s_k]} \hat{T}(s, q_\alpha)(s_i, q_j) -  \hat{T}(s, q_\beta)(s_i, q_j)
    \label{EntryK}
\end{equation}


Two cases arise. First, let us consider the case where $[s] \neq [s_k]$. It follows from the definition of product LMCs that, for every transition from equivalence class $[s]$ to $[s_k]$, there are corresponding transitions $(s, q_\alpha)
\rightarrow (s_k, q_{\alpha'})$ and $(s, q_\beta) \rightarrow (s_k, q_{\beta'})$ (for some $\alpha', \beta'$) in $\mathcal{M}^\times$ such that $\hat{T}(s, q_\alpha)(s_k, q_{\alpha'}) = \hat{T}(s, q_\beta)(s_k,
q_{\beta'}) = T(s)(s_k)$. Thus, the sum equates to 0.


Now, assume $[s] = [s_k]$. Two cases arise. First, let us consider the case where
$[s]$ does not contain a self loop $T^\times (s, q)(s, q') > 0$ for some $q, q'$. Then all terms in the
preceding sum (\ref{EntryK}) are $T(s)(s) = 0$. Now, let us consider the case
where there is such a self loop so that $T(s)(s) > 0$. Then, since the DRA is
deterministic, there is exactly one successor $(s, q') \in [s]$ for any given
$(s, q) \in [s]$. From the pigeonhole principle, any trajectory of such product states resulting from the self loop must inevitably revisit some state, yielding a lollipop walk. For any two states, say $(s, q_i)$
and $(s, q_j)$, in this walk, there is exactly one successor yielding the term
$T(s)(s)$ for each of the two states. Thus, the sum in (\ref{EntryK}) equates
to 0. 

\noindent Since $s, \alpha, \beta$ were chosen arbitrarily, we have shown that $(\mathbf{e}_\alpha
- \mathbf{e}_\beta) \hat{T} V = 0$. 

\section{Proof of Theorem~\ref{theorem:aggregateLMC} and Corollary~\ref{corollary:aggregateIsOriginal}}
\label{sec:new-theorem}
Given an irreducible LMC $\mathcal{M} = (S, \beta, T, R, AP, L)$ and some arbitrary DRA $\mathcal{A}$ with $m$ states, let the corresponding product LMC be given by $\mathcal{M}^\times = (S^\times, \beta^\times, T^\times,$ $ R^\times, AP, L^\times)$. Consider the partition $\bigcup_{s \in S} [s]$ given by the equivalence classes $[s] = \{ (s, q) | (s, q) \in S^\times \}$. We show that $\mathcal{M} = \mathcal{M}^*$, where $\mathcal{M}^* = (S^*, \beta^*, T^*, R^*,$ $ AP, L^*)$ denotes the aggregated LMC of $\mathcal{M}^\times$ under partition $\bigcup_{s \in S} [s]$. From Lemma \ref{ordinarily-lumpable}, we know that this partition is ordinarily lumpable. Let $T([s])([s']) \in [0, 1]^{m \times m}$ denote the transition probability sub-matrix in $\mathcal{M}^\times$ shown below corresponding to transitions from states in $[s]$ to states in $[s']$. Per Theorem \ref{theorem:aggregateLMC}, we have that $T^*(s)(s') = \mathbf{e}_i T([s])([s']) \mathbf{e}^T$. Therefore, we must show that $T(s)(s') = \\ \mathbf{e}_i T([s])([s']) \mathbf{e}^T$, where $i$ is an arbitrary index in $\{i | (s, q_i) \in S^\times \}$.

\begin{equation}
    T([s])([s']) = \left( \begin{matrix} T^\times (s, q_1)(s', q_1) & \cdots & T^\times (s, q_1)(s', q_m) \\ \vdots & \ddots & \vdots \\ T^\times (s, q_m)(s', q_1) & \cdots & T^\times (s, q_m)(s', q_m) \end{matrix} \right) 
\end{equation}
\noindent We have:
\begin{equation}
\begin{aligned}
    \nonumber
    T^*(s)(s') & = \mathbf{e}_i T([s])([s']) \mathbf{e}^T = \left( T^\times (s, q_i)(s', q_k) \right)_k \mathbf{e}^T \\ & = \sum_{k=1}^m T^\times (s, q_i)(s', q_k) = T^\times (s, q_i)(s', q') = T(s)(s')
\end{aligned}
\end{equation}

where $q' = \delta(q_i, L(s'))$ is the transition observed in the DRA. From Theorem \ref{theorem:aggregateLMC}, this establishes that the original LMC $\mathcal{M}$ and the aggregate LMC $\mathcal{M}^*$ have the same transition probability function $T = T^*$. The remaining equivalences follow from Definition \ref{def:quotient} and the definition of product LMCs.

\section{Proof of Lemma~\ref{lemma:key}}
\label{sec:key-lemma}
The multichain transition probability matrix $T^\times$ is given by the canonical form in equation
(\ref{multichainT}) \cite{Puterman:1994}, where $W_k$ denotes transitions from transient states to the $k^\text{th}$ BSCC and $Z$ denotes transitions
between transient states.  
\begin{align}
    T^\times = \begin{bmatrix}
    T_1 & 0 & \ldots & 0  & 0 \\
    0 & T_2 & \dots & 0 & 0 \\
    \vdots & \vdots & \ddots & \vdots & \vdots \\
    0 & 0 & \ldots & T_m &  0 \\
    W_1 & W_2 & \ldots & W_m & Z
\end{bmatrix}
\label{multichainT}
\end{align}
    
    
\noindent Recall the steady-state equations given below for the product LMC $M^\times$.

\begin{equation}
\nonumber
\begin{aligned}
    (\text{Pr}^\infty (s_0, q_0), \dots) T^\times &= (\text{Pr}^\infty (s_0, q_0), \dots) \\ 
    \sum_{(s, q) \in S^\times} \text{Pr}^\infty (s, q) & = 1
\end{aligned}
\end{equation}
    
\noindent This yields the following system of linear equations for $M^\times$:

\begin{equation}
\begin{aligned}
	& \text{Pr}^\infty (s, q) = \!\!\!\! \sum_{(s', q') \in S^k} \!\!\!\! \text{Pr}^\infty (s', q') T^\times (s', q')(s, q) \hspace*{12pt} \forall (s, q) \in S^k, k \\  
    & \sum_{k \leq m} \sum_{(s, q) \in S^k} \!\!\!\! \text{Pr}^\infty (s, q) = 1
\end{aligned}
\label{equation:isomorphicProof1}
\end{equation}
    
\noindent Similarly, we have the following equations for $M^{'}$:

\begin{equation}
\begin{aligned}
& \text{Pr}^\infty (s, q) = \sum_{(s', q') \in S'} \text{Pr}^\infty (s', q') T'(s', q')(s, q) \hspace*{12pt} \forall (s, q) \in S' \\ 
& \sum_{(s, q) \in S'} \!\!\!\! \text{Pr}^\infty (s, q) = 1
\end{aligned}
\label{equation:isomorphicProof2}
\end{equation}

Let $\langle s \rangle = \{(s, q) | (s, q) \text{ isomorphic in } T^G_1, \dots,
T^G_m \}$, where $T^G_k$ denotes the graph structure of $T_k$ in $\mathcal{M}^\times$. See Figure \ref{fig:isomorphic} for an example. It follows from equations (\ref{equation:isomorphicProof1}) and (\ref{equation:isomorphicProof2}) that 
	
\begin{equation}
    \nonumber
    \begin{aligned}
    \sum_{(s, q) \in S'} \!\!\!\! & \text{Pr}^\infty (s, q) = \sum_{(s, q) \in S'} \sum_{(s', q') \in S'} T'(s', q')(s, q) \text{Pr}^\infty (s', q') = 1 \\ & = \sum_{k \leq m} \sum_{(s, q) \in S^k} \!\!\!\! \text{Pr}^\infty (s, q) \\ & = \sum_{k \leq m} \sum_{(s, q) \in S^k} \sum_{(s', q') \in S^k} \!\!\!\! T^\times (s', q')(s, q) \text{Pr}^\infty (s', q') \\ 
    & = \sum_{(s, q) \in S^1} \sum_{(s', q') \in S^1} T^\times (s', q')(s, q) \sum_{(s'', q'') \in \langle s' \rangle} \!\!\!\! \text{Pr}^\infty (s'', q'') \\ 
\end{aligned}
\end{equation}
    
The choice of $S^1$ in the summations of the last equation is arbitrary. Indeed, this result holds for any $S^k$. Therefore, $\text{Pr}^\infty (s, q) = \sum_{(s', q') \in \langle s \rangle} \text{Pr}^\infty (s', q')$ for all $(s, q) \in S'$.

\section{Proof of Lemma~\ref{lemma:key2}}
\label{sec:key2}
	
	Since $\pi(a | s, q) \in \{0, 1\}$ and we would like to prove that $x_{sqa} = \pi(a | s, q) \sum_a x_{sqa}$, we must show that $x_{sqa} \in \{0, \sum_a x_{sqa}\}$ such that $x_{sqa} = \sum_a x_{sqa}$ if and only if $\pi(a | s, q) = 1$. Note that $\sum_a x_{sqa} > 0$ because $(s, q)$ is recurrent. ($\implies$) The contrapositive statement $\pi(a | s, q) = 0$ implies $x_{sqa} = 0$ follows directly from constraint $(iii)$. ($\impliedby$) Since $\pi(a | s, q) = 1$, it follows from constraint $(iv)$ that $\sum_{a' \in A \setminus \{a\}} \pi (a' | s, q) = 0$. From constraint $(iii)$, we then have $\sum_{a' \in A \setminus \{a\}} x_{sqa'} = 0$. It follows that $x_{sqa} = \sum_a x_{sqa}$. From the preceding arguments, we have $x_{sqa} = \pi(a | s, q) \text{Pr}^\infty_\pi (s, q)$. Since the product LMC has BSCCs with some shared state in $S$, it follows from Lemma \ref{lemma:multiToUni} that the original LMC $\mathcal{M}_\pi$ will be a unichain. Thus, the solution to the steady-state equations (\ref{steadyStateEquations}) for $\mathcal{M}_\pi$ will be the unique stationary distribution reflecting the true steady-state behavior of the agent. We further have that this solution to equations (\ref{steadyStateEquations}) for $\mathcal{M}_\pi$ are obtained from $x$ in constraints $(i)$ and $(ii)$ per Lemma \ref{lemma:key}. Thus, $\pi(a | s, q) = x_{sqa} / \sum_a x_{sqa} \in \{0, 1\}$ is the policy which yields the steady-state distribution given by the solution to constraints $(i)$ through $(iv)$. 
	
	\section{Proof of Theorem~\ref{thm:main}}
    \label{sec:theorem-main}
	($\implies$) We assume that $s_0$ is the initial state in the LMDP $\mathcal{M}$ (i.e. $\beta(s_0) = 1)$. Since the solution $(x, f, \pi, \mathcal{I})$ is feasible, there must be some $(s, q) \in S \times \bigcup_i G_i$ such that constraint $(xi)$ is satisfied. From constraint $(ix)$, note that this is only possible if $\mathcal{I}_{sq} = 1$, which is only possible if there is incoming flow into $(s, q)$ per the flow variables $f_{s'q'sq}$ in constraint $(vi)$. This corresponds to the outgoing flow of some neighboring state $(s', q')$. Again, from constraint $(vi)$, it follows that $(s', q')$ must also have incoming flow. By induction, we see that this holds for all states leading from the initial state $(s_0, \delta(q_0, L(s_0)))$ to $(s, q)$. Since positive flow is only enabled along edges  corresponding to actions chosen by the solution policy $\pi$ (per constraint $(v)$), it follows that there is a path from $(s_0, \delta(q_0, L(s_0)))$ to $(s, q)$ in the product LMC $(\mathcal{M} \times \mathcal{A})_\pi$ and, due to the one-to-one correspondence between paths in the product LMC and the original LMC $\mathcal{M}_\pi$, there is also a path from $s_0$ to $s$ in the latter. Now, consider the policy defined at state $(s, q)$. Since there is incoming flow, it follows from constraint $(viii)$ that there must also be outgoing flow. This flow will continue from state to state until some state is revisited, creating a BSCC. Constraints $(xii)$ through $(xvi)$ ensure that all such BSCCs contain the same states. In particular, note that the satisfaction of constraint $(xi)$ yields $\mathcal{I}^k = 1$ for the $k^\text{th}$ AMEC (without loss of generality) per constraint $(xii)$. Similarly, since $\mathcal{I}_{sq} = 1$, it follows from constraint $(xiv)$ that $\mathcal{I}^k_s = 1$ for that same AMEC. Constraint $(xiii)$ ensures that $\mathcal{I}^k_s$ cannot be set to $1$ unless there is some $\mathcal{I}_{sq} = 1$ in the $k^\text{th}$ AMEC. Thus, for any state $s \in S$ and $k$, we have $\mathcal{I}^k_s \leq \mathcal{I}^k$ and this is only met with equality if state $s \in S$ shows up in the BSCC corresponding to the $k^\text{th}$ AMEC. It follows that the sum $\sum_k (\mathcal{I}^k_s - \mathcal{I}^k)$ on the RHS of constraint $(xv)$ achieves its maximum value of $0$ if and only if state $s$ shows up in every BSCC of the product LMC $(\mathcal{M} \times \mathcal{A})_\pi$. This must be the case per constraint $(xvi)$, which forces the LHS of constraint $(xv)$ to be $0$ for some state. From Lemma \ref{lemma:multiToUni}, this yields a unichain original LMC $\mathcal{M}_\pi$ in the original LMDP $\mathcal{M}$. From Theorem \ref{theorem:aggregateLMC}, Corollary \ref{corollary:aggregateIsOriginal}, and Lemma \ref{lemma:key}, the satisfaction of constraints $(i)$, $(ii)$, and $(x)$ yields an LMC $\mathcal{M}_\pi$ satisfying the steady-state specifications in $\theta$. Finally, the satisfaction of constraints $(iii)$ and $(iv)$ ensure that the solution policy $\pi$ is deterministic.
	
		
	($\impliedby$) Assume we have a deterministic policy $\pi$ and the unichain original LMC induced by $\pi$ is given by $\mathcal{M}_\pi = (S, T_\pi)$ and satisfies the given SS+LTL specification $\theta$. We derive a feasible solution $(x, f, \pi, \mathcal{I})$ to program (\ref{equation:finalProgramss}). It follows from Lemma \ref{lemma:multiToUni} that all BSCCs in the product LMC $(\mathcal{M} \times \mathcal{A})_\pi$ obtained from $\mathcal{M}_\pi$ and the DRA $\mathcal{A}$ of $\phi_\text{LTL}$ must contain the same states in $S$. 
	Let $\mathcal{I}_{sq} = \mathcal{I}_s = \mathcal{I}^k = \mathcal{I}^k_s = 1$ for all states $(s, q)$ in the BSCCs of $(\mathcal{M} \times \mathcal{A})_\pi$ (these correspond to AMECs in the product LMDP $\mathcal{M} \times \mathcal{A}$). This satisfies constraints $(xii)$ through $(xvi)$. Note that the steady-state probabilities of such states must be positive and we can assign these values to $\sum_a x_{sqa}$ for every $(s, q)$ in the BSCCs. This satisfies $(i), (ii)$, and $(ix)$. Since the SS+LTL specification is satisfied, it must be the case that some state $(s, q) \in S \times \bigcup_i G_i$ is visited infinitely often, so we can assign the steady-state probability of such states to $\sum_a x_{sqa}$, thereby satisfying $(xi)$. Furthermore, since the steady-state operators in the specification are satisfied, it follows that constraint $(x)$ is satisfied. Note that constraints $(v)$ through $(viii)$ can be satisfied by setting the flow values to be proportional to the policy values. Finally, since $\pi(a | s, q) = x_{sqa} / \sum_a x_{sqa} \in \{0, 1\}$ ($\pi(a | s, q) = 0$) for recurrent (transient) states per Lemma \ref{lemma:key2}, constraint $(iii)$ is satisfied and constraint $(iv)$ follows since deterministic policies are probability distributions.

\clearpage

\section{Figure}
\label{appendix:figure}

\begin{figure}[h]
  \centering
  \includegraphics[width=8cm]{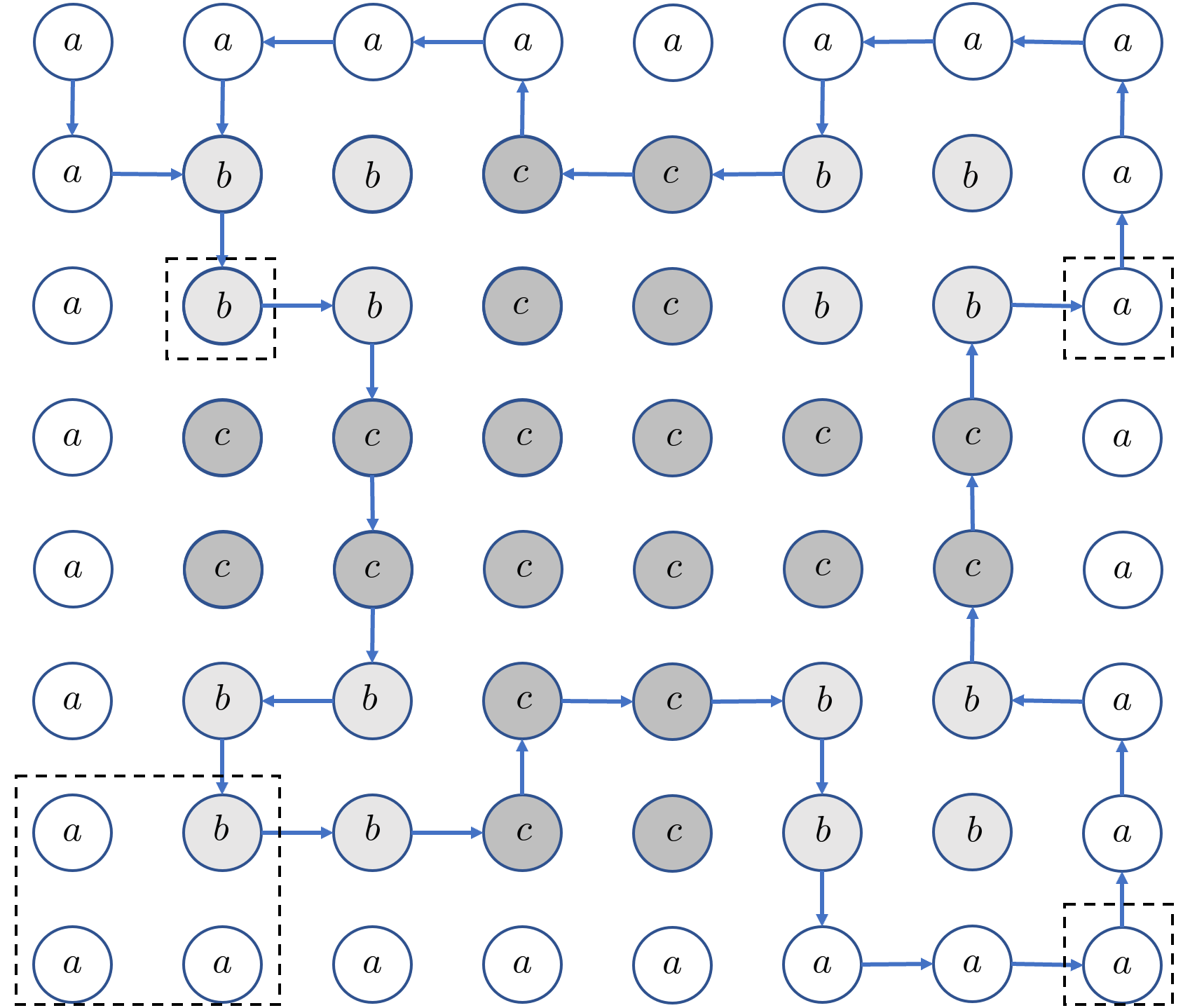}
  \caption{Let $s_{ij}$ denote the state in row $i$, column $j$ of the above $8 \times 8$ grid with deterministic transition dynamics defined in the obvious way. Let the SS+LTL specification be defined by the LTL specification $\mathbf{G F } (a \wedge \mathbf{X } b \wedge \mathbf{X X } c \wedge \mathbf{X X X } c)$ and four SS specifications, each defined over the dashed subsets of states. All of these regions must be visited at least $0.01$ proportion of the time. The solution policy found using program (\ref{equation:finalProgramss}) is denoted by the arrows. Note that the induced original LMC is a unichain (with two transient states). Furthermore, it satisfies the given SS+LTL specification. Indeed, the LTL specification can be visually verified and the steady-state probabilities for each of the dashed subsets of states is $0.0294$.}
  \label{fig:programExample}
\end{figure}

\end{document}


\linenumbers  


\section*{Appendix A: Related Work}

The set of all maximal end-component for an LMDP can be efficiently computed~\cite{Chatte11}

The controller synthesis problem given $\omega$-regular objectives has been studied at length in the literature, particularly under the name of LTL controller synthesis \cite{etessami2007multi,yannakakis2008multi,forejt2011automated}. The problem of deriving a control policy which satisfies constraints on the steady-state distribution of the resulting agent has been studied more recently \cite{SSC,ijcai2019-784,ijcai2020-563}. However, the literature on solving expected-reward constrained MDPs has often studied similar problems given that the expected-reward objective leverages the steady-state distribution or occupation measures, which are analogous to the steady-state distribution over state-action pairs, in order to determine expected policy values \cite{kallenberg1983linear,Puterman:1994,altman_total_cost_98,4927531}. However, to the best of the authors' knowledge, these two facets of asymptotic behavior have not yet been incorporated for the controller synthesis problem.

LTL controller synthesis approaches generally begin by computing the product LMDP from the original LMDP and the automaton representation of the given objective. Then, the union of accepting maximal end components (AMECs) are computed and a policy is found such that the agent reaches some such component. Once there, actions can be chosen arbitrarily such that all states within the AMEC are visited infinitely often, thereby ensuring that the acceptance condition of the DRA is met and the objective is therefore satisfied by said policy. Generally, this choice of actions within the AMECs is arbitrary. However, it is evident that these choices are critical for our proposed logic. Indeed, these choices affect the steady-state distribution of the resulting agent as determined by the LMC induced by the solution policy. 

In more restricted LMDP problems, such as those found under the umbrella of Stochastic Shortest Path (SSP) problems, flow-based programs have been proposed to reach goal states, sometimes under the presence of constraints \cite{iDual}, \cite{iDual2}, \cite{occupationMeasurePlanning}, \cite{occupationPLTL}. While most methods for solving constrained MDPs revolve around the use of mathematical programs, some reinforcement learning approaches have also been proposed for optimizing the average-reward objective and, to a lesser extent, for solving constrained instances of average-reward MDPs. Some noteworthy examples include the constrained actor-critic method proposed in \cite{ConstrainedActorCritic_12}, wherein a Lagrangian relaxation of the problem is used to incorporate steady-state costs into the objective function being optimized by the constrained actor-critic algorithm. A similar Lagrangian Q-learning approach is proposed in \cite{QLearningConstrained_13}. Both of these reinforcement learning methods assume that every Markov chain induced by a policy is irreducible.

Perhaps the work most relevant to the results established in this paper stems from \cite{SSC} and \cite{ijcai2019-784}. In \cite{SSC}, the Steady-State Control (SSC) problem is introduced. This is then generalized as Steady-State Policy Synthesis (SSPS) in \cite{ijcai2019-784}. In particular, the SSC problem entails finding a policy whose induced Markov chain satisfies a given steady-state distribution. This problem assumes that the underlying MDP is ergodic in that every policy yields recurrent Markov chains. This ensures that steady-state distributions reflect the true asymptotic behavior of the Markov chain. This is a fairly common assumption. Indeed, as observed recently by Altman in \cite{Altman2019_11} with regards to average-reward or -cost problems in constrained MDPs, ``\textit{The existing theory for solving such problems requires strong assumptions on the ergodic structure of the problem}.'' In that work, a linear program is proposed to find optimal stochastic policies to constrained MDP formulations without relying on the ergodic assumption. In \cite{ijcai2019-784}, the SSPS problem is posed as a generalization of SSC by allowing steady-state constraints to contain inequalities as well as probability intervals. The solution proposed therein does not assume ergodic MDPs and instead finds a strongly connected Markov chain within an arbitrary MDP, if one exists, such that steady-state constraints are satisfied. However, that approach cannot handle transient states nor multichain MDPs. These issues were addressed in \cite{ijcai2020-563}, wherein a solution to the steady-state planning problem is proposed for multichain MDPs by focusing on a restricted class of policies, such as imposing that all actions be taken with some probability by the solution policy.

Our proposed logic and controller synthesis solution unify much of the foregoing by reasoning about both $\omega$-regular properties as well as steady-state distributions simultaneously. The proposed approach finds an optimal expected-reward control policy that satisfies a given SSTL specification and does not make strong assumptions on the ergodicity nor connectivity of the underlying LMDP.

\section*{Appendix B: Figures}

\begin{figure}
	\includegraphics[width=\columnwidth]{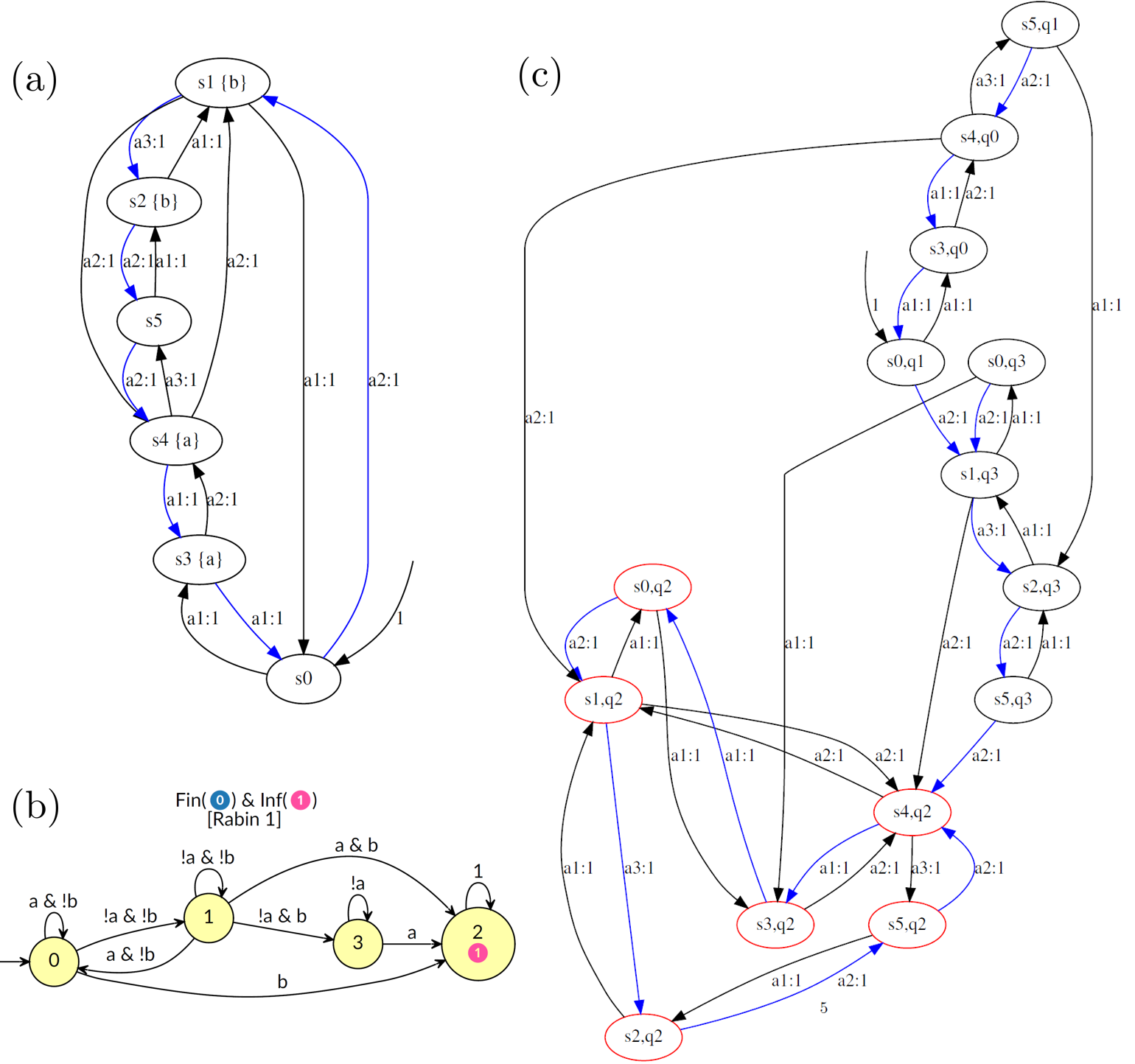}
	\caption{(a) LMDP $\mathcal{M} = (S, \beta, A, T, R = \emptyset, AP, L)$, where $S = \{s_0, \dots, s_5\}$, $\beta(s_0) = 1$, $A = \{a_1, a_2, a_3\}$, $AP = \{a, b\}$, and $L(s_1) = L(s_2) = \{b\}, L(s_3) = L(s_4) = \{a\}$. The transition function is deterministic (i.e. $T: S \times A \times S \mapsto \{0, 1\}$) and shown in the figure by the transitions $a_i:1$ between states $s, s'$ denoting that $T(s' | s, a_i) = 1$. The blue transitions define a policy $\pi: S \times A \mapsto \{0, 1\}$ which induces a recurrent Markov chain $\mathcal{M}_\pi$ visiting all states in $\mathcal{M}$. (b) Given the LTL specification $(\textbf{F } a) \textbf{ U } b$, the LTL DRA $\mathcal{A} = (Q, q_0, \Sigma, \delta, F = \{(J_i, K_i)\}_i)$ is defined, where $Q = \{q_0, q_1, q_2, q_3\}$, $\Sigma = 2^{AP}$, $F = \{(\emptyset, \{q_2\})\}$, and the transition function is given by $\delta(q_0, \{a\}) = q_0, \delta(q_0, \emptyset) = q_1, \delta(q_0, \{b\}) = q_2, \delta(q_1, \{a\}) = q_0, \delta(q_1, \emptyset) = q_1, \delta(q_1, \{b\}) = q_3, \delta(q_3, \{a\}) = q_2, \delta(q_3, \{b\}), \delta(q_3, \{a\}) = q_2, \delta(q_2, \cdot) = q_2$. (c) Product LMDP $\mathcal{M} \times \mathcal{A}$, where red nodes represent states in the accepting MEC of $\mathcal{M} \times \mathcal{A}$. The blue transitions define the LMC $(\mathcal{M} \times \mathcal{A})_\pi$ induced by the policy $\pi$ given by the blue transitions in (a) for the LMDP $\mathcal{M}$. Note that this policy induces an LMC that has probability 1 of being trapped in the accepting MEC. Therefore, the probability of satisfying $(\textbf{F } a) \textbf{ U } b$ given that we start in state $s_0$ is 1.}
	\label{fig:example}
\end{figure}

\begin{figure}
	\includegraphics[width=\columnwidth]{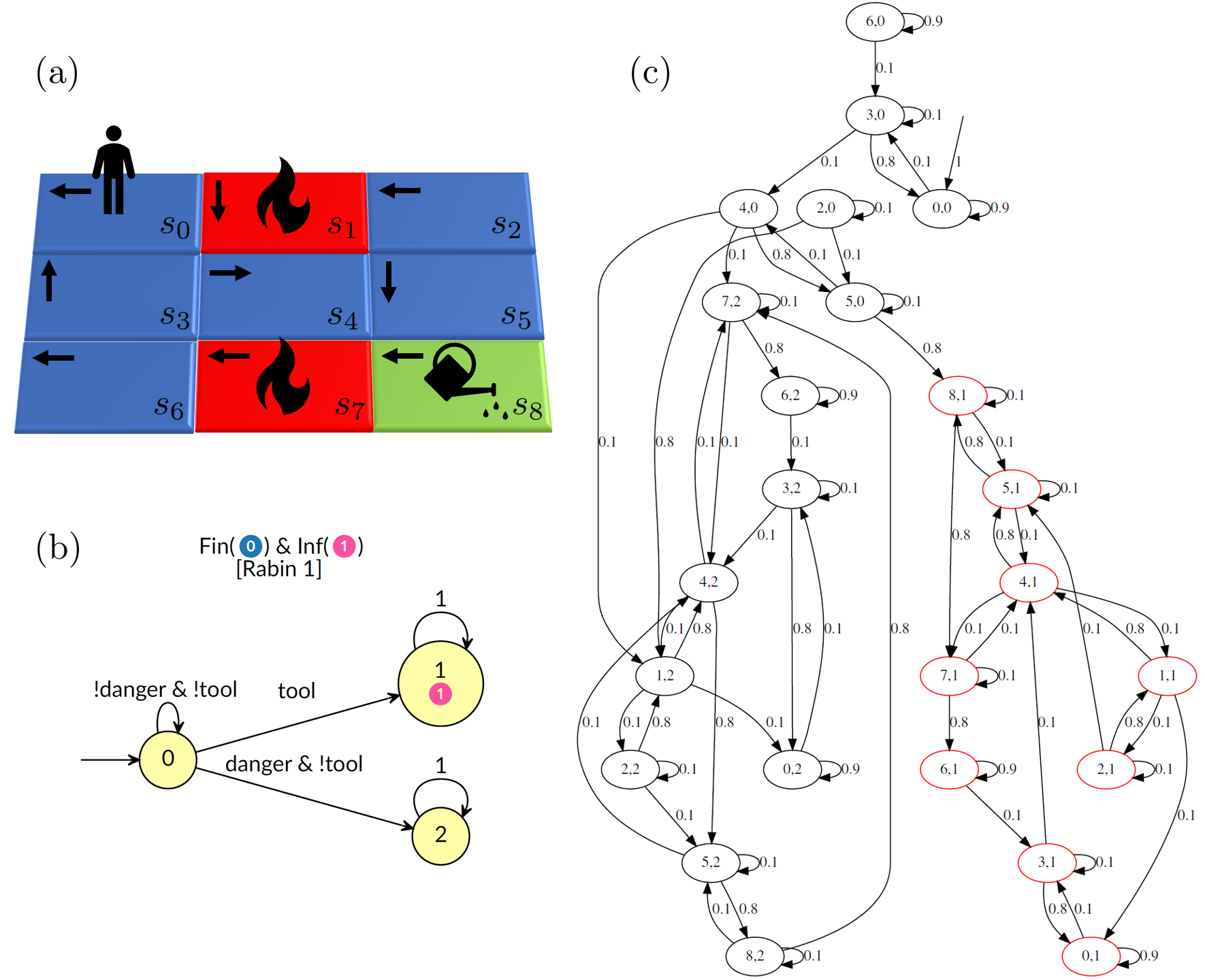}
	\caption{(a) LMDP $\mathcal{M} = (S, \beta, A, T, R = \emptyset, AP, L)$, where $S = \{s_0, \dots, s_8\}$, $\beta(s_0) = 1$, $A = \{\leftarrow, \downarrow, \rightarrow, \uparrow\}$, $AP = \{home, danger, tool\}$, and $L(s_0) = \{home\}, L(s_1) = L(s_7) = \{danger\}, L(s_8) = \{tool\}$. the agent has a chance of slipping whenever it moves, causing a transition into one of three possible states. If the agent chooses to go right (left), there is an 80$\%$ chance that it will transition to the right (left), and the chance of transitioning to either of the states above or below it is 10$\%$. Similarly, if the agent chooses to up (down), it will end up in the states above (below) it with 80$\%$ chance, and in the states to the right and left of it with chance 10$\%$ each. In the corners of the map, the agent may stay in place with 90$\%$ probability by choosing to move against the boundary of the map (e.g. $T(s_0 | s_0, \leftarrow) = 0.9$). (b) Given the specification $(! danger \textbf{U} tool) \wedge \textbf{SS}_{[0.75, 1]} home$, the LTL DRA $\mathcal{A} = (Q, q_0, \Sigma, \delta, F = \{(J_i, K_i)\}_i)$ is defined, where $Q = \{q_0, q_1, q_2\}$, $\Sigma = 2^{AP}$, $F = \{(\emptyset, \{q_1\})\}$, and the transition function is given by $\delta(q_0, \{home\}) = q_0, \delta(q_0, \{tool\}) = q_1, \delta(q_0, \{danger\}) = q_2, \delta(q_1, \cdot) = q_1, \delta(q_2, \cdot) = q_2$. The steady-state operator is not used in defining $\mathcal{A}$. (c) Product LMC $(\mathcal{M} \times \mathcal{A})_\pi$ induced by the policy $\pi$ given by the black arrows in (a) for the product LMDP $\mathcal{M} \times \mathcal{A}$, where red nodes represent states in the accepting BSCC of $(\mathcal{M} \times \mathcal{A})_\pi$. Note that this policy has non-zero probability of being trapped in the accepting BSCC. Furthermore, note that $\sum_{s \in \mathcal{L}_{home}} \text{Pr}^\infty_\pi (s) = \text{Pr}^\infty_\pi (s_0) = 0.76$, thereby satisfying the steady-state operator $\textbf{SS}_{[0.75, 1]} home$.}
	\label{fig:example2}
\end{figure}

\section*{Appendix C: Proofs}

\subsection{Proof of Lemma 1}

\begin{proof}
We must show that, for all $s \in \hat{S}$ and all $(s, q_\alpha), (s,
q_\beta) \in [s]$, we have $(\mathbf{e}_\alpha - \mathbf{e}_\beta) \hat{T} V = 0$. Let $V \in \{0,
1\}^{(|S||Q|) \times |S|}$ such that $v_{(ij)k} = 1$ if $(s_i, q_j) \in
[s_k]$ and $v_{(ij)k} = 0$ otherwise. 
\begin{equation}
\begin{aligned}
    (\mathbf{e}_\alpha - \mathbf{e}_\beta) \hat{T} V = \left( \hat{T}(s, q_\alpha)(s_i, q_j) -  \hat{T}(s, q_\beta)(s_i, q_j) \right)_{\substack{i \leq |S| \\ j \leq |Q|}} V\\ 
    = \left( \sum_{ij} v_{(ij)k} [\hat{T}(s, q_\alpha)(s_i, q_j) -  \hat{T}(s, q_\beta)(s_i, q_j)] \right)_{k \leq |S|}
\end{aligned}
\end{equation}

\noindent Looking at the $k^\text{th}$ entry, we have:

\begin{equation}
    \sum_{(s_i, q_j) \in [s_k]} \hat{T}(s, q_\alpha)(s_i, q_j) -  \hat{T}(s, q_\beta)(s_i, q_j)
    \label{EntryK}
\end{equation}


Two cases arise. First, let us consider the case where $[s] \neq [s_k]$. It follows that, for every transition from equivalence class $[s]$ to $[s_k]$, there are corresponding transitions $(s, q_\alpha)
\rightarrow (s_k, q_{\alpha'})$ and $(s, q_\beta) \rightarrow (s_k, q_{\beta'})$ in $M^\times$ such that $\hat{T}(s, q_\alpha)(s_k, q_{\alpha'}) = \hat{T}(s, q_\beta)(s_k,
q_{\beta'}) = T(s)(s_k)$. Thus, the sum equates to 0.


Now, assume $[s] = [s_k]$. Two cases arise. First, let us consider the case where
$[s]$ does not contain a self loop $T^\times (s, q)(s, q') > 0$ for some $q, q'$. Then all terms in the
preceding sum (\ref{EntryK}) are $T(s)(s) = 0$. Now, let us consider the case
where there is such a self loop so that $T(s)(s) > 0$. Then, since the DRA is
deterministic, there is exactly one successor $(s, q') \in [s]$ for any given
$(s, q) \in [s]$. From the pigeonhole principle, any trajectory of such product states resulting from the self loop must inevitably revisit some state, yielding a lollipop walk. For any two states, say $(s, q_i)$
and $(s, q_j)$, in this walk, there is exactly one successor yielding the term
$T(s)(s)$ for each of the two states. Thus, the sum in (\ref{EntryK}) equates
to 0. 

\noindent Since $s, \alpha, \beta$ were chosen arbitrarily, we have shown that $(\mathbf{e}_\alpha
- \mathbf{e}_\beta) \hat{T} V = 0$. \qed
\end{proof}

\subsection*{Proof of Corollary 1}

\begin{proof}
Given an irreducible LMC $M = (S, \beta, T, R, AP, L)$ and some arbitrary DRA $A$ with $m$ states, let the corresponding product LMC be given by $M^\times = (S^\times, \beta^\times, T^\times, R^\times, AP, L^\times)$. Consider the partition $\bigcup_{s \in S} [s]$ given by the equivalence classes $[s] = \{ (s, q) | (s, q) \in S^\times \}$ and let $M' = (S', \beta', T', R', AP, L')$ denote the quotient LMC induced by this partition on $M^\times$. We show that $M = M^*$, where $M^* = (S^*, \beta^*, T^*, R^*, AP, L^*)$ denotes the aggregated LMC of $M'$. Let $T([s])([s']) \in [0, 1]^{m \times m}$ denote the transition probability sub-matrix in $M^\times$ corresponding to transitions from states in $[s]$ to states in $[s']$. Per Theorem \ref{theorem:aggregateLMC}, we have that $T^*(s)(s') = \mathbf{e}_i T([s])([s']) \mathbf{e}^T$. Therefore, we must show that $T(s)(s') = \mathbf{e}_i T([s])([s']) \mathbf{e}^T$, where $i$ is an arbitrary index in $\{i | (s, q_i) \in S^\times \}$.

\begin{equation}
    T([s])([s']) = \left( \begin{matrix} T^\times (s, q_1)(s', q_1) & \cdots & T^\times (s, q_1)(s', q_m) \\ \vdots & \ddots & \vdots \\ T^\times (s, q_m)(s', q_1) & \cdots & T^\times (s, q_m)(s', q_m) \end{matrix} \right) 
\end{equation}
\noindent We have:
\begin{equation}
\begin{aligned}
    \nonumber
    T^*(s)(s') & = \mathbf{e}_i T([s])([s']) \mathbf{e}^T = \left( T^\times (s, q_i)(s', q_k) \right)_k \mathbf{e}^T \\ & = \sum_{k=1}^m T^\times (s, q_i)(s', q_k) = T^\times (s, q_i)(s', q') = T(s)(s')
\end{aligned}
\end{equation}

where $q' = \delta(q_i, L(s'))$ is the transition observed in the DRA. From Theorem \ref{theorem:aggregateLMC}, this establishes that the original LMC $M$ and the aggregate LMC $M^*$ have the same transition probability function $T = T^*$. The remaining equivalences follow from Definition \ref{def:quotient} and the definition of product LMCs. \qed
\end{proof}

\subsection*{Proof of Lemma 2}

\begin{proof}
Both directions follow straightforwardly from the one-to-one correspondence between paths in $M$ and paths in $M^\times$. Furthermore, its single BSCC $S' \subseteq S$ of $M$ is such that $S' = \{s | (s, q) \in \bigcup_k S^k\}$. \qed
\end{proof}

\subsection*{Proof of Lemma 3}

\begin{proof}
The multichain transition probability matrix $T$ is given by the canonical form in equation
(\ref{multichainT})~\cite{Puterman:1994}, where $W_k$ denotes transitions
from transient states to the $k^\text{th}$ BSCC and $Z$ denotes transitions
between transient states.  
\begin{align}
    T = \begin{bmatrix}
    T_1 & 0 & \ldots & 0  & 0 \\
    0 & T_2 & \dots & 0 & 0 \\
    \vdots & \vdots & \ddots & \vdots & \vdots \\
    0 & 0 & \ldots & T_m &  0 \\
    W_1 & W_2 & \ldots & W_m & Z
\end{bmatrix}
\label{multichainT}
\end{align}
    
    
\noindent Recall the steady-state equations given below for the product LMC $M$.

\begin{equation}
\nonumber
\begin{aligned}
    (\text{Pr}^\infty (s_0, q_0), \dots) T &= (\text{Pr}^\infty (s_0, q_0), \dots) \\ 
    \sum_{(s, q) \in S^\times} \text{Pr}^\infty (s, q) & = 1
\end{aligned}
\end{equation}
    
\noindent This yields the following system of linear equations for $M$:

\begin{equation}
\begin{aligned}
	& \text{Pr}^\infty (s, q) = \!\!\!\! \sum_{(s', q') \in S^k} \!\!\!\! \text{Pr}^\infty (s', q') T(s', q')(s, q) \hspace*{12pt} \forall (s, q) \in S^k, k \\  
    & \sum_{k \leq m} \sum_{(s, q) \in S^k} \!\!\!\! \text{Pr}^\infty (s, q) = 1
\end{aligned}
\label{equation:isomorphicProof1}
\end{equation}
    
\noindent Similarly, we have the following equations for $M^*$:

\begin{equation}
\begin{aligned}
& \text{Pr}^\infty (s, q) = \sum_{(s', q') \in S^*} \text{Pr}^\infty (s', q') T(s', q')(s, q) \hspace*{12pt} \forall (s, q) \in S^* \\ 
& \sum_{(s, q) \in S^*} \!\!\!\! \text{Pr}^\infty (s, q) = 1
\end{aligned}
\label{equation:isomorphicProof2}
\end{equation}

Let $\langle s \rangle = \{(s, q) | (s, q) \text{ isomorphic in } T^G_1, \dots,
T^G_m \}$, where $T^G_k$ denotes the graph structure of $T_k$ in $M$. It follows from equations (\ref{equation:isomorphicProof1}) and (\ref{equation:isomorphicProof2}) that 
	
\begin{equation}
    \nonumber
    \begin{aligned}
    \sum_{(s, q) \in S^*} \!\!\!\! & \text{Pr}^\infty (s, q) = \sum_{(s, q) \in S^*} \sum_{(s', q') \in S^*} T(s', q')(s, q) \text{Pr}^\infty (s', q') = 1 \\ & = \sum_{k \leq m} \sum_{(s, q) \in S^k} \!\!\!\! \text{Pr}^\infty (s, q) = \sum_{k \leq m} \sum_{(s, q) \in S^k} \sum_{(s', q') \in S^k} \!\!\!\! \text{Pr}^\infty (s', q') T(s', q')(s, q) \\ 
    & = \sum_{(s, q) \in S^1} \sum_{(s', q') \in S^1} T(s', q')(s, q) \sum_{(s'', q'') \in \langle s' \rangle} \!\!\!\! \text{Pr}^\infty (s'', q'') \\ 
\end{aligned}
\end{equation}
    
Note that the choice of $S^1$ in the last equation is arbitrary. Therefore,
$\text{Pr}^\infty (s, q) = \sum_{(s', q') \in \langle s \rangle}
\text{Pr}^\infty (s', q')$ for all $(s, q) \in S^*$. \qed
\end{proof}

\subsection*{Proof of Lemma 4}

\begin{proof}
	
	Since $\pi(a | s, q) \in \{0, 1\}$ and $x_{sqa}$ must equal $\pi(a | s, q) \sum_a
	x_{sqa}$, we must show that $x_{sqa} \in \{0, \sum_a x_{sqa}\}$ such that
	$x_{sqa} = \sum_a x_{sqa}$ if and only if $\pi(a | s, q) = 1$. Note that $\sum_a x_{sqa} > 0$ because $(s, q)$ is
	recurrent.  
	
	($\implies$) The contrapositive statement $\pi(a | s, q) = 0$ implies $x_{sqa}
	= 0$ follows directly from constraint $(iii)$.  
	
	($\impliedby$) Since $\pi(a | s, q) = 1$, it follows from constraint $(iv)$
	that $\sum_{a' \in A \setminus \{a\}} \pi (a' | s, q) = 0$. From constraint
	$(iii)$, we then have $\sum_{a' \in A \setminus \{a\}} x_{sqa'} = 0$. It
	follows that $x_{sqa} = \sum_a x_{sqa}$. 
	
	From the preceding arguments, we have $x_{sqa} = \pi(a | s, q)
	\text{Pr}^\infty_\pi (s, q)$. From the unichain condition, we further have
	that there is a unique solution to constraints $(i)$ and $(ii)$. Thus, $\pi$
	is the policy which yields the steady-state distribution given by the
	solution to constraints $(i)$ through $(iv)$. \qed
\end{proof}

\subsection*{Proof of Theorem 2}

\begin{proof}
	($\implies$) We assume that $s_0$ is the initial state in the LMDP (i.e. $\beta(s_0) = 1)$. Since the solution $(x, f, \pi, \mathcal{I})$ is feasible,
	there must be some $(s, q) \in S \times \bigcup_i G_i$ such that constraint
	$(xi)$ is satisfied. From constraint $(ix)$, note that this is only possible
	if $\mathcal{I}_{sq} = 1$, which is only possible if there is incoming flow
	into $(s, q)$ per the flow variables $f_{s'q'sq}$ in constraint $(vi)$. This
	corresponds to the outgoing flow of some neighboring state $(s', q')$.
	Again, from constraint $(vi)$, it follows that $(s', q')$ must also have
	incoming flow. By induction, we see that this holds for all states leading
	from the initial state $(s_0, \delta(q_0, L(s_0)))$ to $(s, q)$. Since
	positive flow is only enabled along edges  corresponding to actions chosen
	by the solution policy $\pi$ (per constraint $(v)$), it follows that there
	is a path from $(s_0, \delta(q_0, L(s_0)))$ to $(s, q)$ in the product LMC
	and, due to the one-to-one correspondence between paths in the product LMC
	the original LMC, there is also a path from $s_0$ to $s$ in the latter. Now,
	consider the policy defined at state $(s, q)$. Since there is incoming flow,
	it follows from constraint $(viii)$ that there must also be outgoing flow.
	This flow will continue from state to state until some state is revisited,
	creating a BSCC. Constraints $(xii)$ through $(xvi)$ ensure that all such
	BSCCs contain the same states. In particular, note that the satisfaction of
	constraint $(xi)$ yields $\mathcal{I}^k = 1$ for the $k^\text{th}$ AMEC
	(without loss of generality) per
	constraint $(xii)$. Similarly, since $\mathcal{I}_{sq} = 1$, it follows from
	constraint $(xiv)$ that $\mathcal{I}^k_s = 1$ for that same AMEC. Constraint
	$(xiii)$ ensures that $\mathcal{I}^k_s$ cannot be set to $1$ unless there is
	some $\mathcal{I}_{sq} = 1$ in the $k^\text{th}$ AMEC. Thus, for any state $s \in S$ and $k$, we have
	$\mathcal{I}^k_s \leq \mathcal{I}^k$ and this is only met with equality if
	state $s \in S$ shows up in the BSCC corresponding to the $k^\text{th}$ AMEC. It follows that the sum $\sum_k
	(\mathcal{I}^k_s - \mathcal{I}^k)$ on the RHS of constraint $(xv)$ achieves
	its maximum value of $0$ if and only if state $s$ shows up in every BSCC of
	the product LMC. This must be the case per constraint $(xvi)$, which forces
	the LHS of constraint $(xv)$ to be $0$ for some state. From Lemma 2, this
	yields a unichain in the original LMDP. From Theorem 2 and Lemma 3, the
	satisfaction of constraint $(x)$ yields an LMC satisfying 
	steady-state specifications in $\theta$. 
	
		
	($\impliedby$) Assume we have a policy $\pi$ and the unichain induced by
	$\pi$ is given by $\mathcal{M}_\pi = (S, T_\pi)$ and satisfies the given
	SSTL specification. We derive a feasible solution $(x, f, \pi,
	\mathcal{I})$ to program (\ref{equation:finalProgramss}). It
	follows from Lemma 2 that all BSCCs in the product LMDP $\mathcal{M}^\times_\pi$ must
	contain the same states in $S$. 
	Let $\mathcal{I}_{sq} = \mathcal{I}_s = \mathcal{I}^k =
	\mathcal{I}^k_s = 1$ for all states $(s, q)$ in the BSCCs of $M^\times_\pi$ (these correspond to AMECs in $M^\times$). This satisfies constraints $(xii)$ through
	$(xvi)$. Note that the steady-state probabilities of such states must be
	positive and we can assign these values to $\sum_a x_{sqa}$ for every $(s,
	q)$ in the BSCCs. This satisfies $(i), (ii)$, and $(ix)$. Since the
	SSTL specification is satisfied, it must be the case that some state
	$(s, q) \in S \times \bigcup_i G_i$ is visited infinitely often, so we can assign the steady-state probability of such states to $\sum_a x_{sqa}$, thereby satisfying $(xi)$. Furthermore, since the
	steady-state operators in the specification are satisfied, it follows that constraint $(x)$ is satisfied. Note that constraints $(v)$
	through $(viii)$ can be satisfied by setting the flow values to be
	proportional to the policy values. 
	\end{proof}

\section*{Appendix D: Verbose Integer Linear Program}

\begin{figure*}[h!]
	\begin{gather*}
	\nonumber
	\begin{aligned}
	\max & \sum_{(s, q) \in S^\times} \sum_{a \in A(s)} x_{sqa} \sum_{s' \in S} T(s' | s, a) R(s, a, s') \hspace*{12pt} \text{subject to} \\ 
	(i) \text{ } & \sum_{(s, q) \in S^\times} \sum_{a \in A(s)} x_{sqa} T^\times((s', q') | (s, q), a) = \sum_{a \in A(s')} x_{s'q'a} & \hspace*{12pt} \forall (s', q') \in S^\times \\ 
	(ii) \text{ } & \sum_{(s, q) \in S^\times} \sum_{a \in A(s)} x_{sqa} = 1 \\ 
	(iii) \text{ } & x_{sqa} \leq \pi(a | s) & \forall (s, q) \in S^\times, a \in A \\ 
	(iv) \text{ } & \sum_{a \in A(s)} \pi \left(a | s \right) = 1 & \forall s \in S \\ 
	(v) \text{ } & f_{sqs'q'} \leq \sum_{a \in A(s)} T((s', q') | (s, q), a) \pi(a | s) & \hspace*{12pt} \forall ((s, q), (s', q')) \in T^G \\ 
	(vi) \text{ } & \sum_{((s', q'), (s, q)) \in T^G} f_{s'q'sq} \geq \sum_{((s, q), (s', q')) \in T^G} f_{sqs'q'} + \epsilon \mathcal{I}_{sq} & \hspace*{12pt} \forall (s, q) \in S^\times \setminus \{(s_0, \delta(q_0, L(s_0)))\} \\ 
	(vii) \text{ } & \sum_{((s', q'), (s, q)) \in T^G} f_{s'q'sq} \leq \mathcal{I}_{sq} & \hspace*{12pt} \forall (s, q) \in S^\times \\ 
	(viii) \text{ } & \sum_{((s, q), (s', q')) \in T^G} f_{sqs'q'} \geq \sum_{((s', q'), (s, q)) \in T^G} f_{s'q'sq} / 2 & \hspace*{12pt} \forall (s, q) \in S^\times \\ 
	(ix) \text{ } & \sum_{a \in A(s)} x_{sqa} \leq \mathcal{I}_{sq} & \forall (s, q) \in S^\times \\ 
	(x) \text{ } & l \leq \sum_{s \in L^{-1}(\psi)} \sum_{q \in Q} \sum_{a \in A(s)} x_{sqa} \leq u & \forall \textbf{SS}_{[l, u]} \psi \in \theta \\ 
	(xi) \text{ } & \sum_{s \in S} \sum_{q \in \bigcup_i K_i} \sum_{a \in A(s, q)} x_{sqa} > 0 \\ 
	(xii) \text{ } & \sum_{(s, q) \in AMEC_k} \sum_a x_{sqa} \leq \mathcal{I}^k & \forall k \\ 
	(xiii) \text{ } & \mathcal{I}^k_s \leq \sum_{(s, q) \in AMEC_k} \mathcal{I}_{sq} & \forall s \in S, k \\ 
	(xiv) \text{ } & \sum_{(s, q) \in AMEC_k} \frac{\mathcal{I}_{sq}}{|Q|} \leq \mathcal{I}^k_s & \forall s \in S, k \\ 
	(xv) \text{ } & \mathcal{I}_s - 1 \leq \frac{\sum_k \left( \mathcal{I}^k_s - \mathcal{I}^k \right)}{\# of AMECs} & \forall s \in S \\ 
	(xvi) \text{ } & \sum_s \mathcal{I}_s \geq 1 \\ 
	& x_{sqa}, f_{sqs'q'} \in [0, 1] & \hspace*{-88pt} \forall ((s, q), a, (s', q')) \in S \times A \times S^\times \\ 
	& \mathcal{I}_{sq}, \mathcal{I}_s, \mathcal{I}^k, \mathcal{I}^k_s \in \{0, 1\} & \hspace*{-168pt} \forall ((s, q), a) \in S^\times \times A \\ 
	\end{aligned}
	\label{equation:finalProgram}
	\end{gather*}
\end{figure*}

The notion of quotient and aggregate LMCs is used in some proofs. The definition follows. Some theorems are referenced from other work and thus do not require a proof. They are nonetheless listed here for completeness.

\begin{definition}[Quotient and Aggregate Markov Chain]
    Given a product LMC $M = (S^\times, \beta^\times, T^\times, R^\times, AP, L^\times)$ and its partition into equivalence classes $[s] = \{(s, q) | (s, q) \in S^\times\}$, the corresponding quotient LMC is given by $M' = (S', \beta', T', R', AP, L')$, where $S' = \{[s_i]\}_i$, $\beta'([s]) = \sum_{(s, q) \in [s]} \beta^\times (s, q)$, $T'([s'] | [s]) = T^\times ((s', \cdot) | (s, \cdot))$, $R'([s], [s']) = R^\times ((s, \cdot), (s', \cdot))$, and $L'([s]) = L^\times ((s, \cdot))$. The aggregated LMC associated with $M'$ is given by $M* = (S^*, \beta^*, T^*, R^*, AP, L^*)$ is defined in the obvious way (i.e. $S^* = \{s | [s] \in S'\}, \beta^* (s) = \beta' ([s]), T^* (s' | s) = T' ([s'] | [s]), R^* (s, s') = R' ([s], [s']), L^* (s) = L'([s])$).
    \label{def:quotient}
\end{definition}

\begin{lemma}
Given an arbitrary BSCC $(\hat{S}, \hat{T})$ of a product LMC $M^\times =
(S^\times, T^\times_\pi)$, the partition $\bigcup_{s \in S} [s]$ given by equivalence classes $[s] =
\{(s, q) | (s, q) \in \hat{S}\}$ is ordinarily lumpable. 
\end{lemma}

\begin{proof}
We must show that, for all $s \in \hat{S}$ and all $(s, q_\alpha), (s,
q_\beta) \in [s]$, we have $(e_\alpha - e_\beta) \hat{T} V = 0$. Let $V \in \{0,
1\}^{|S| \times |Q| \times |S|}$ such that $v_{ijk} = 1$ if $(s_i, q_j) \in
[s_k]$ and $v_{ijk} = 0$ otherwise. 
\begin{equation}
\begin{aligned}
    (e_\alpha - e_\beta) \hat{T} V = \left( \hat{T}((s, q_k) | (s, q_\alpha)) -  \hat{T}((s, q_k) | (s, q_\beta)) \right)_k V\\ 
    = \left( \sum_{ij} v_{ijk} [\hat{T}((s_i, q_j) | (s, q_\alpha)) -  \hat{T}((s_i, q_j) | (s, q_\beta))] \right)_k
\end{aligned}
\end{equation}

Looking at the $k^\text{th}$ entry, we have:

\begin{equation}
    \sum_{(s_i, q_j) \in [s_k]} \hat{T}((s_i, q_j) | (s, q_\alpha)) -  \hat{T}((s_i, q_j) | (s, q_\beta))
    \label{EntryK}
\end{equation}


Two cases arise. First, let us consider the case where $s \neq s_k$. By
definition of the quotient LMC, it follows that, for every transition $[s]
\rightarrow [s_k]$, there are corresponding transitions $(s, q_\alpha)
\rightarrow (s_k, q_{\alpha'})$ and $(s, q_\beta) \rightarrow (s_k, q_{\beta'})$
such that $\hat{T}((s_k, q_{\alpha'}) | (s, q_\alpha)) = \hat{T}((s_k,
q_{\beta'}) | (s, q_\beta)) = T(s_k | s)$. Thus, the sum equates to 0 in this
case. 


Now, assume $s = s_k$. Two cases arise. First, let us consider the case where
$[s]$ does not contain a self loop in the quotient LMC. Then all terms in the
preceding sum (\ref{EntryK}) are $T(s | s) = 0$. Now, let us consider the case
where there is a self loop so that $T(s | s) > 0$. Then, since the DRA is
deterministic, there is exactly one successor $(s, q') \in [s]$ for any given
$(s, q) \in [s]$. From the pigeonhole principle, any trajectory of such product states resulting from the self loop must inevitably revisit some state, yielding a lollipop walk. For any two states, say $(s, q_i)$
and $(s, q_j)$, in this walk, there is exactly one successor yielding the term
$T(s | s)$ for each of the two states. Thus, the sum in (\ref{EntryK}) equates
to 0. 

Since $s, \alpha, \beta$ were chosen arbitrarily, we have shown that $(e_\alpha
- e_\beta) \hat{T} V = 0$. 
\end{proof}

\begin{theorem}[\cite{lumpabilityOriginal}, \cite{lumpability}, Theorem 4]
Given a Markov chain $M' = (S', T')$ and an ordinarily lumpable partition
$\bigcup_{s' \in S'} [s']$ of $S'$, the steady-state distribution of the
aggregated Markov chain $M = (S, T)$ satisfies $\text{Pr}^\infty (s) = \sum_{s'
\in [s]} \text{Pr}^\infty (s')$ for every $s \in S$. Furthermore, the transition
function of the aggregated MC is given by $T(s' | s) = \mathbf{e}_i T([s'] |
[s]) \mathbf{e}^T$, where $i$ is an arbitrary index in the set $\{i | (s, q_i) \in S^\times \}$.
\label{theorem:aggregateLMC}
\end{theorem}

\begin{corollary}
Given an irreducible product LMC, the original LMC is the aggregated MC
resulting from the ordinarily lumpable partition. 
\end{corollary}

\begin{proof}
Given an irreducible LMC $M = (S, \beta, T, R, AP, L)$ and some arbitrary DRA $A$ with $m$ nodes, let the corresponding product LMC be given by $M^\times = (S^\times, \beta^\times, T^\times, R^\times, AP, L^\times)$. Consider the partition $\bigcup_{s \in S} [s]$ given by the equivalence classes $[s] = \{ (s, q) | (s, q) \in S^\times \}$ and let $M' = (S', \beta', T', R', AP, L')$ denote the quotient LMC induced by this partition on $M^\times$. We show that $M = M^*$, where $M^* = (S^*, \beta^*, T^*, R^*, AP, L^*)$ denotes the aggregate LMC of $M'$. Let $T([s'] | [s]) \in [0, 1]^{m \times m}$ denote the transition probability sub-matrix in $M^\times$ corresponding to transitions from states in $[s]$ to states in $[s']$. Per Theorem \ref{theorem:aggregateLMC}, we have that $T^*(s' | s) = \mathbf{e}_i T([s'] | [s]) \mathbf{e}^T$. Therefore, we must show that $T(s' | s) = \mathbf{e}_i T([s'] | [s]) \mathbf{e}^T$.

\begin{equation}
    T([s'] | [s]) = \left( \begin{smallmatrix} T^\times ((s', q_1) | (s, q_1)) & \cdots & T^\times ((s', q_m) | (s, q_1)) \\ \vdots & \ddots & \vdots \\ T^\times ((s', q_1) | (s, q_m)) & \cdots & T^\times ((s', q_m) | (s, q_m)) \end{smallmatrix} \right) 
\end{equation}

\noindent We have:

\begin{equation}
\begin{aligned}
    \nonumber
    T^*(s' | s) & = \mathbf{e}_i T([s'] | [s]) \mathbf{e}^T \\ & = \left( T^\times ((s', q_k) | (s, q_i)) \right)_k \mathbf{e}^T \\ & = \sum_{k=1}^m T^\times ((s', q_k) | (s, q_i)) \\ & = T^\times ((s', q') | (s, q_i)) = T(s' | s)
\end{aligned}
\end{equation}

where $q' = \delta(q_i, L(s'))$ is the transition observed in the DRA. From Theorem \ref{theorem:aggregateLMC}, this establishes that the original LMC $M$ and the aggregate LMC $M^*$ have the same transition probability function $T = T^*$. The remaining equivalences follow from Definition \ref{def:quotient} and the definition of product LMCs.
\end{proof}

The previous theorem and corollary establish the one-to-one correspondence
between the steady-state probabilities derived for a unichain LMC and the
steady-state distribution for the original LMC. Now, let us consider the case
where the product LMC is a multichain. We establish sufficient conditions for
establishing the same one-to-one correspondence of steady-state distributions. 

\begin{lemma}
Let $\pi$ denote a policy and $M_\pi = (S, T_\pi)$ denote the Markov chain
induced by this policy such that $M_\pi$ satisfies $\phi_\text{LTL}$. Let
$(S^k)_k$ denote the BSCCs in the product MC induced by $\pi$. Then $M_\pi$ is a
unichain if and only if some state $(s, \cdot) \in S^\times$ shows up in every
BSCC $S^k$ of the product LMC. 
\end{lemma}

\begin{proof}
(Sketch) ($\implies$) Assume $M_\pi$ is a unichain. From the one-to-one
correspondence of paths between the LMC and product LMC, it follows that all
BSCCs in the product LMC must be the same. 
($\impliedby$) If some state is shared across all BSCCs in the product LMC,
then, by the one-to-one correspondence between path in $M_\pi$ and paths in
$M^\times_\pi$, it follows that $M_\pi$ is irreducible. Furthermore, its single
BSCC $S'$ is such that $S' = S^k$ for all $k$. 
\end{proof}

\begin{lemma}
    Given a multichain $M = (S, T)$ with $m$ identical BSCCs given by transition
    probability matrices $T_1, \dots, T_m$ (That is, the graph structures of
    these components are all isomorphic) and an irreducible Markov chain $M^* =
    (S^*, T^*)$, where $S^*$ contains exactly the states in the first BSCC and
    $T^* = T_1$ (without loss of generality), the steady-state probability of an
    arbitrary state $(s, q) \in S^*$ is equivalent to the sum of steady-state
    probabilities of all states isomorphic to it in $S$.  
\end{lemma}
    
\begin{proof}
The transition probability matrix $T$ is given by the canonical form in equation
(\ref{multichainT})~\cite{Puterman:1994}, where $W_k$ denotes transitions
from transient states to the $k^\text{th}$ BSCC and $Z$ denotes transitions
between transient states.  
\begin{align}
    T = \begin{bmatrix}
    T_1 & 0 & \ldots & 0  & 0 \\
    0 & T_2 & \dots & 0 & 0 \\
    \vdots & \vdots & \ddots & \vdots & \vdots \\
    0 & 0 & \ldots & T_m &  0 \\
    W_1 & W_2 & \ldots & W_m & Z
\end{bmatrix}
\label{multichainT}
\end{align}
    
    
\noindent Recall the steady-state equations given below for the product LMC $M$.

\begin{equation}
\nonumber
\begin{aligned}
    (\text{Pr}^\infty (s_0, q_0), \dots) T &= (\text{Pr}^\infty (s_0, q_0), \dots) \\ 
    \sum_{(s, q) \in S^\times} \text{Pr}^\infty (s, q) & = 1
\end{aligned}
\end{equation}
    
\noindent This yields the following system of linear equations for $M$:

\begin{equation}
\nonumber    
\begin{aligned}
	& \text{Pr}^\infty (s, q) = \!\!\!\! \sum_{(s', q') \in S^k} \!\!\!\! \text{Pr}^\infty (s', q') T(s, q | s', q') \hspace*{12pt} \forall (s, q) \in S^k, k \\  
    & \sum_{k \in [m]} \sum_{(s, q) \in S^k} \!\!\!\! \text{Pr}^\infty (s, q) = 1
\end{aligned}
\end{equation}
    
\noindent Similarly, we have the following equations for $M^*$:

\begin{equation}
\nonumber    
\begin{aligned}
& \text{Pr}^\infty (s, q) = \sum_{(s', q') \in S^*} \text{Pr}^\infty (s', q') T(s, q | s', q') \hspace*{12pt} \forall (s, q) \in S^* \\ 
& \sum_{(s, q) \in S^*} \!\!\!\! \text{Pr}^\infty (s, q) = 1
\end{aligned}
\end{equation}

Let $\langle s \rangle = \{(s, q) | (s, q) \text{ isomorphic in } T^G_1, \dots,
T^G_m \}$, where $T^G_k$ denotes the graph structure of $T_k$ in $M$. It follows that 
	
\begin{equation}
    \nonumber
    \begin{aligned}
    & \sum_{(s, q) \in S^*} \!\!\!\! \text{Pr}^\infty (s, q) = \\ & \sum_{(s, q) \in S^*} \sum_{(s', q') \in S^*} T(s, q | s', q') \text{Pr}^\infty (s', q') = 1 = \\ & \sum_{k \in [m]} \sum_{(s, q) \in S^k} \!\!\!\! \text{Pr}^\infty (s, q) = \\ 
    & \sum_{k \in [m]} \sum_{(s, q) \in S^k} \sum_{(s', q') \in S^k} \!\!\!\! \text{Pr}^\infty (s', q') T(s, q | s', q') = \\ 
    & \sum_{(s, q) \in S^1} \sum_{(s', q') \in S^1} T(s, q | s', q') \sum_{(s'', q'') \in \langle s' \rangle} \!\!\!\! \text{Pr}^\infty (s'', q'') \\ 
\end{aligned}
\end{equation}
    
Note that the choice of $S^1$ in the last equation is arbitrary. Therefore,
$\text{Pr}^\infty (s, q) = \sum_{(s', q') \in \langle s \rangle}
\text{Pr}^\infty (s', q')$ for all $(s, q) \in S^*$.
\end{proof}

\begin{lemma}
	Let $(x, \pi)$ denote a feasible solution to constraints $(i)$ through
	$(iv)$ and assume that the Markov chain $M_\pi$ induced by $\pi$ is
	unichain. Then $x_{sqa} = \pi (a | s) \text{Pr}^\infty_\pi (s, q) = \pi (a |
	s) \sum_a x_{sqa}$ for all recurrent states $(s, q) \in S^\times$. 
\end{lemma}

\begin{proof}
	
	Since $\pi(a | s) \in \{0, 1\}$ and $x_{sqa}$ must equal $\pi(a | s) \sum_a
	x_{sqa}$, we must show that $x_{sqa} \in \{0, \sum_a x_{sqa}\}$ such that
	$x_{sqa} = \sum_a x_{sqa}$ if and only if $\pi(a | s) = 1$. By
	contrapositive, this also establishes that $x_{sa} = 0$ if and only if
	$\pi(a | s) = 0$. Note that $\sum_a x_{sqa} > 0$ because $(s, q)$ is
	recurrent.  
	
	($\implies$) The contrapositive statement $\pi(a | s) = 0$ implies $x_{sqa}
	= 0$ follows directly from constraint $(iii)$.  
	
	($\impliedby$) Since $\pi(a | s) = 1$, it follows from constraint $(iv)$
	that $\sum_{a' \in A \setminus \{a\}} \pi (a' | s) = 0$. From constraint
	$(iii)$, we then have $\sum_{a' \in A \setminus \{a\}} x_{sqa'} = 0$. It
	follows that $x_{sqa} = \sum_a x_{sqa}$. 
	
	From the preceding arguments, we have $x_{sqa} = \pi(a | s)
	\text{Pr}^\infty_\pi (s, q)$. From the unichain condition, we further have
	that there is a unique solution to constraints $(i)$ and $(ii)$. Thus, $\pi$
	is the policy which yields the steady-state distribution given by the
	solution to constraints $(i)$ through $(iv)$. 
\end{proof}

\begin{theorem}
    Given an LMDP $\mathcal{M} = (S, \beta, A, T, R, AP, L)$ and an SSTL
	objective $\theta$
	let $(x, f, \pi, \mathcal{I})$ denote an assignment to the variables in
	program (11). Then 
    $(x, f, \pi, \mathcal{I})$ is a feasible solution if and only if
    $\mathcal{M}_\pi = (S_\pi, \beta, T_\pi, AP, L)$ satisfies $\theta$ and is a
    unichain. 
\end{theorem}

\begin{proof}
	($\implies$) We assume that $s_0$ is the initial state in the LMDP (i.e. $\beta(s_0) = 1)$. Since the solution $(x, f, \pi, \mathcal{I})$ is feasible,
	there must be some $(s, q) \in S \times \bigcup_i K_i$ such that constraint
	$(xi)$ is satisfied. From constraint $(ix)$, note that this is only possible
	if $\mathcal{I}_{sq} = 1$, which is only possible if there is incoming flow
	into $(s, q)$ per the flow variables $f_{s'q'sq}$ in constraint $(vi)$. This
	corresponds to the outgoing flow of some neighboring state $(s', q')$.
	Again, from constraint $(viii)$, it follows that $(s', q')$ must also have
	incoming flow. By induction, we see that this holds for all states leading
	from the initial state $(s_0, \delta(q_0, L(s_0)))$ to $(s, q)$. Since
	positive flow is only enabled along edges  corresponding to actions chosen
	by the solution policy $\pi$ (per constraint $(v)$), it follows that there
	is a path from $(s_0, \delta(q_0, L(s_0)))$ to $(s, q)$ in the product LMC
	and, due to the one-to-one correspondence between paths in the product LMC
	the original LMC, there is also a path from $s_0$ to $s$ in the latter. Now,
	consider the policy defined at state $(s, q)$. Since there is incoming flow,
	it follows from constraint $(viii)$ that there must also be outgoing flow.
	This flow will continue from state to state until some state is revisited,
	creating a BSCC. Constraints $(xii)$ through $(xvi)$ ensure that all such
	BSCCs contain the same states. In particular, note that the satisfaction of
	constraint $(xi)$ yields $\mathcal{I}^k = 1$ for the $k^\text{th}$ AMEC
	(without loss of generality) within which some state $(s, q)$ resides per
	constraint $(xii)$. Similarly, since $\mathcal{I}_{sq} = 1$, it follows from
	constraint $(xiv)$ that $\mathcal{I}^k_s = 1$ for that same AMEC. Constraint
	$(xiii)$ ensures that $\mathcal{I}^k_s$ cannot be set to $1$ unless there is
	some $\mathcal{I}_{sq} = 1$. Thus, for any state $s \in S$ and $k$, we have
	$\mathcal{I}^k_s \leq \mathcal{I}^k$ and this is only met with equality if
	state $s \in S$ shows up in BSCC $k$. It follows that the sum $\sum_k
	(\mathcal{I}^k_s - \mathcal{I}^k)$ on the RHS of constraint $(xv)$ achieves
	its maximum value of $0$ if and only if state $s$ shows up in every BSCC of
	the product LMC. This must be the case per constraint $(xvi)$, which forces
	the LHS of constraint $(xv)$ to be $0$ for some state. From Lemma 2, this
	yields a unichain in the original LMDP. From Theorem 3 and Lemma 3, the
	satisfaction of constraint $(x)$ yields an LMC which satisfies the
	steady-state specifications in $\theta$. 
	
		
	($\impliedby$) Assume we have a policy $\pi$ and the unichain induced by
	$\pi$ is given by $\mathcal{M}_\pi = (S, T_\pi)$ and satisfies the given
	SSTL specification. We derive a feasible solution $(x, f, \pi,
	\mathcal{I})$ to program (11) in the sequel. It
	follows from Lemma 2 that all BSCCs in $(\mathcal{M} \times A)_\pi$ must
	contain the same states in $S$. That is, $S^k = S^{k'}$ for all BSCCs in the
	product LMC. Let $\mathcal{I}_{sq} = \mathcal{I}_s = \mathcal{I}^k =
	\mathcal{I}^k_s = 1$ for all states $(s, q)$ in the BSCC corresponding to
	enabled AMEC $k$ (for all $k$). This satisfies constraints $(xii)$ through
	$(xvi)$. Note that the steady-state probabilities of such states must be
	positive and we can assign these values to $\sum_a x_{sqa}$ for every $(s,
	q)$ in the BSCCs. This satisfies $(i), (ii)$, and $(ix)$. Since the
	SSTL specification is satisfied, it must be the case that some state
	$(s, q) \in S \times \bigcup_i K_i$ is visited infinitely often, so we can
	assign the steady-state probability of such states to the sum of variables
	$\sum_a x_{sqa}$, thereby satisfying $(xi)$. Furthermore, since the
	steady-state operators in the specification are satisfied, it follows from
	Theorem 3 that constraint $(x)$ is satisfied. Note that constraints $(v)$
	through $(viii)$ can be satisfied by setting the flow values to be
	proportional to the policy values. 
\end{proof}

\newpage
\clearpage

\bibliography{refs}